\documentclass[amssymb,aps,prd,nofootinbib,floatfix,superscriptaddress,secnumarabic,preprintnumbers]{revtex4}

\usepackage{graphicx}
\usepackage{slashed}
\usepackage{amsmath}

\usepackage{xcolor}
\usepackage{caption}
\usepackage{subcaption}
\usepackage{comment}
\usepackage{tikz}
\usetikzlibrary{positioning}

\newcommand{\be}{\begin{equation}}
\newcommand{\ee}{\end{equation}}
\newcommand{\bea}{\begin{eqnarray}}
\newcommand{\eea}{\end{eqnarray}}

\newcommand{\MSbar}{{\overline{\rm MS}}}
\newcommand{\GIRS}{{\rm GIRS}}
\newcommand{\LR}{{\rm L}}
\newcommand{\DR}{{\rm DR}}
\newcommand{\Tr}{{\rm Tr}}
\newcommand{\gtilde}{\frac{g^2}{16 \, \pi^2}\; }

\begin{document}
\vspace*{1cm}

\title{Nonperturbative renormalization of the supercurrent in \texorpdfstring{$\cal{N}$}{N} = 1 Supersymmetric Yang-Mills Theory \texorpdfstring{\bigskip}{}}

\author{G.~Bergner}
\email[]{georg.bergner@uni-jena.de}
\affiliation{University of Jena, Institute for Theoretical Physics, \\
  Max-Wien-Platz 1, D-07743 Jena, Germany\bigskip}

\author{M.~Costa}
\email[]{kosta.marios@ucy.ac.cy}
\affiliation{Department of Physics, University of Cyprus,  \\
  P.O. Box 20537, 1678 Nicosia, Cyprus\bigskip}
\affiliation{Department of Chemical Engineering, Cyprus University of Technology, \\
30 Archbishop Kyprianou Str., 3036, Limassol, Cyprus\bigskip}  

\author{H.~Panagopoulos}
\email[]{haris@ucy.ac.cy}
\affiliation{Department of Physics, University of Cyprus,  \\
  P.O. Box 20537, 1678 Nicosia, Cyprus\bigskip}

\author{S.~Piemonte\smallskip}
\email[]{stefano.piemonte@ur.de}
\affiliation{University of Regensburg, Institute for Theoretical Physics, \\
  Universit\"{a}tsstr. 31, D-93040 Regensburg, Germany\bigskip}


\author{I.~Soler Calero}
\email[]{ivan.soler.calero@uni-jena.de}
\affiliation{University of Jena, Institute for Theoretical Physics, \\
  Max-Wien-Platz 1, D-07743 Jena, Germany\bigskip}

\author{G.~Spanoudes\bigskip}
\email[]{g.spanoudis@cyi.ac.cy}
\affiliation{Computation-based Science and Technology Research Center, \\
  The Cyprus Institute, 20 Constantinou Kavafi Str., 2121 Nicosia, Cyprus\bigskip}

\begin{abstract}
  In this work, we study the nonperturbative renormalization of the supercurrent operator in $\mathcal{N} = 1$ Supersymmetric Yang-Mills (SYM) theory, using a gauge-invariant renormalization scheme (GIRS). The proposed prescription addresses successfully the unwanted mixing of the supercurrent with other operators of equal or lower dimension, which respect the same global symmetries. This mixing is introduced by the unavoidable breaking of supersymmetry on the lattice. In GIRS all gauge-noninvariant operators, which mix with the supercurrent, are excluded from the renormalization procedure. The one remaining mixing operator is accessible by numerical simulations. We present results for the renormalization of the supercurrent using a GIRS scheme. We also compute at one-loop order the conversion matrix which relates the nonperturbative renormalization factors in GIRS to the reference scheme $\MSbar$.
  
\end{abstract}

\maketitle

\section{Introduction}
\label{Introduction}
Supersymmetric (SUSY) theories have long been considered a promising completion of the standard model with appealing properties like natural explanations for dark matter. Unbroken supersymmetry implies that particles arrange themselves into multiplets  with the same number of fermionic and bosonic degrees of freedom. The supersymmetric partners of the Standard Model (SM) particles have not been observed in experiments so far. An unknown breaking mechanism is therefore required in a realistic extension of the Standard Model. In order to understand possible breaking scenarios, it is essential to investigate the nonperturbative regime of SUSY theories. 

Another important motivation for nonperturbative investigations of SUSY theories are theoretical conjectures about the confinement mechanism and relations to Gauge/Gravity duality. These have their foundations in the enhanced symmetries of SUSY gauge theories and it would be interesting to extend and relate them to QCD or Yang-Mills theory. This requires more general insights into the nonperturbative regime of SUSY theories.

Numerical lattice simulations would be an ideal general nonperturbative first principles tool to investigate SUSY gauge theories. However, it is unavoidable to break SUSY in any non-trivial theory on the lattice.
In general, fine tuning is required to restore supersymmetry in the continuum limit, which can be guided by signals provided by the SUSY Ward identities. The analysis of SUSY Ward identities requires the renormalization of the supercurrent, which mixes due to broken supersymmetry with other operators of the same or lower dimension. In our current studies, we investigate how to determine the mixing in a perturbative and nonperturbative way. The first step is a study of $\mathcal{N}=1$ supersymmetric Yang-Mills theory (SYM).

In  Ref.~\cite{curci_supersymmetry_1987} Curci and Veneziano have shown that only a tuning of the gluino mass is required to restore supersymmetry in the continuum limit with Wilson fermions. There is even a very basic connection between chiral symmetry and supersymmetry in SYM, as shown in \cite{Suzuki:2012pc}.  The theory has been successfully simulated using this approach and the results show consistency with supersymmetry \cite{Bergner:2015adz,Ali:2019agk}, where the low energy spectrum was obtained in the continuum limit. The tuning has been done based on chiral symmetry, confirming a consistency with supersymmetric Ward identities (WI) afterwards \cite{farchioni_supersymmetric_2002, Ali:2018fbq,Ali:2020mvj}.  On the lattice the renormalized WI get corrections due to mixing of the supercurrent with all possible operators of the same or lower dimension that respect all unbroken symmetries on the lattice and share the same quantum numbers. Mixing also accounts for an additive renormalization of the fermion mass \cite{curci_supersymmetry_1987}. The fermion mass renormalization provides a signal for the gluino mass tuning. The mixing of the supercurrent appears as an additional undetermined parameter. In the standard approach, mixing and mass renormalization have to be determined simultaneously from the same set of numerically determined WI. In previous investigations, the mixing has also been determined to one loop order in perturbation theory \cite{Taniguchi:1999fc}.

This approach, despite being successful for SYM, has limitations when applied to more general SUSY gauge theories. The number of tuning parameters and mixing coefficients is significantly larger in this case. Therefore it is essential to find alternative ways to determine the renormalization of the supercurrent and reduce the number of parameters that need to be determined from the WI. In this work, we explore an alternative way of renormalizing the supercurrent on the lattice, using a gauge-invariant renormalization scheme (GIRS).

GIRS has been employed in recent studies of operator mixing regarding the renormalization of the trace-less gluon and quark energy-momentum tensor operators in QCD~\cite{Costa:2021iyv} and the gluino-glue operator in SYM theory~\cite{Costa:2021pfu}. In the spirit of X-space scheme~\cite{Gimenez:2004me, Chetyrkin:2010dx, Cichy:2012is, Tomii:2018zix}, GIRS involves Green's functions of two or more gauge-invariant operators in coordinate space. The main advantage of this scheme is that all gauge-noninvariant operators appearing in the set of mixing operators are automatically excluded from the renormalization procedure since their Green's functions are zero. In order to set appropriate conditions in GIRS, which are applicable in both continuum and lattice, we first perform a perturbative calculation of certain Green's functions involving all the gauge-invariant mixing operators in a continuum regularization. Our final goal is to renormalize the supercurrent in the $\MSbar$ scheme, which is the typical scheme used in the analysis of experimental data. Since $\MSbar$ is defined perturbatively in dimensional regularization (DR), we employ DR and we extract the one-loop conversion matrix between GIRS and $\MSbar$. The renormalization factors and mixing coefficients of the supercurrent in $\MSbar$ on the lattice will be extracted at the end by combining our results for the perturbative conversion matrix with the nonperturbative GIRS mixing matrix on the lattice.

The paper is organized as follows: Sec.~\ref{Formulation} contains the formulation of our calculation including the $\mathcal{N}=1$ SYM action, the set of operators under study, as well as an introduction to the GIRS renormalization prescription and to the Green's functions calculated in this work. In Sec.~\ref{Perturbative calculation}, we present our perturbative calculation in dimensional regularization and we provide one-loop results for the Green's functions, the renormalization factors/mixing coefficients, and the conversion matrix between the GIRS and $\MSbar$ schemes. A description of the lattice setup along with the nonperturbative results for the renormalization of the supercurrent is presented in Sec.~\ref{Nonperturbative calculation}. Finally, we conclude in Sec.~\ref{Discussion} with a summary and a discussion of our results and possible future extensions of our work.

 \newpage

\section{GIRS in supersymmetric Yang-Mills}
\label{Formulation}

In this section, we introduce the setup of our calculation. We provide details on the action, the operators and the Green's functions that we calculate in this work.

\subsection{The action of supersymmetric Yang-Mills theory}

In our study, we consider $\mathcal{N} = 1$ supersymmetric Yang-Mills (SYM) theory with gauge group SU($N_c$) in the continuum and on the lattice. By applying the Wess-Zumino gauge~\cite{Wess:1992cp} and by eliminating auxiliary fields of the theory, we end up with the following gauge-fixed continuum action in Euclidean space
\be 
\mathcal{S}_{\rm SYM}= \int d^4 x \left[ \frac{1}{4}u_{\mu \nu}^{\alpha} (x) u_{\mu \nu}^{\alpha} (x) + \frac{1}{2} \bar{\lambda}^{\alpha} (x) \gamma_{\mu} \mathcal{D}_{\mu} \lambda^{\alpha} (x) - \frac{1}{2 \xi} (\partial_\mu u_\mu^{\alpha} (x))^2 - \bar{c}^{\alpha} (x) \partial_\mu D_\mu^{\alpha \beta} c^{\beta} (x) \right],
\label{susylagr}
\ee
where $u_\mu^{\alpha} (x)$ ($\lambda^{\alpha} (x)$, $c^{\alpha} (x)$) is the gluon (gluino, ghost) field, $u_{\mu \nu}^{\alpha}$ is the gluon field-strength tensor and $\xi$ is the gauge parameter [$\xi = 1(0)$ corresponds to Feynman (Landau) gauge]. The gluino field is a Majorana fermion, and thus, $\lambda$ and $\bar{\lambda}$ are related through a charge conjugation transformation. Since supersymmetry is broken in all known regularizations (including dimensional and lattice regularizations) at intermediate steps, we choose the gauge-fixing and ghost terms arising from the Faddeev-Popov procedure to be the same as in the nonsupersymmetric case. Due to the gauge fixing, the total action is no longer gauge-invariant but it is Becchi-Rouet-Stora-Tyutin (BRST) invariant.      
On the lattice, we employ a tree-level Symanzik improved gauge action~\cite{Symanzik:1983dc} and Wilson/clover fermions~\cite{Sheikholeslami:1985ij} for the gluino fields; the action reads
\bea
    {\cal S}^{L}_{\rm SYM}=a^{4}\sum_{x}&\Bigg\{& \frac{2}{g^2} \left[ \frac{5}{3} \sum_{\rm plaq.} {\rm Re} \ {\rm tr}_c (1- U_{\rm plaq.}) - \frac{1}{12} \sum_{\rm rect.} {\rm Re} \ {\rm tr}_c (1-U_{\rm rect.}) \right] \nonumber \\
 &+& \sum_{\mu}\left[ {\rm tr}_c (\bar\lambda\gamma_{\mu} D_{\mu} \lambda) -\frac{a r}{2} {\rm tr}_c (\bar\lambda D^{2}\lambda)\right] - \sum_{\mu, \nu}\left(\frac{c_{\rm SW} \ a}{4}\bar\lambda^{\alpha}\sigma_{\mu \nu}\hat{F}_{\mu \nu}^{\alpha \beta}\lambda^{\beta}\right) + m_0 {\rm tr}_c (\bar\lambda \lambda)\Bigg\},
\label{susylagrLattice}
\eea
where $r \ (c_{\rm SW})$ is the Wilson (clover) parameter, $\sigma_{\mu \nu}=\frac{1}{2} [\gamma_{\mu},\gamma_{\nu}]$, $U_{\rm plaq.} (U_{\rm rect.})$ denotes $1 {\times} 1$ ($2 {\times} 1$) rectangular Wilson loops, $\hat{F}_{\mu \nu}^{\alpha \beta}$ is the clover definition of the field strength tensor in the adjoint representation and $m_0$ is the Lagrangian mass. The explicit definitions of $U_{\rm plaq.}$, $U_{\rm rect.}$, $\hat{F}_{\mu \nu}^{\alpha \beta}$ in terms of link variables $U_\mu (x)$, and the covariant derivatives are standard; they can be found in, e.g.\, Ref.~\cite{Costa:2020keq}. Note that gauge-fixing and ghost terms must be added in the above action, as well as a ``measure'' term coming from the change of integration variables $U_\mu \to u_\mu$. These terms are chosen to be equal to the corresponding terms in the nonsupersymmetric case (see, e.g.~\cite{Costa:2017rht}).

\subsection{Supercurrent and mixing operators}

In standard notation, the supercurrent~\cite{deWit:1975veh} is defined as\footnote{For ease of notation, we leave out the one free Dirac index in all operators appearing in the text. Similarly, we drop the two Dirac indices from all the Green's functions appearing in the sequel.}:
\be
S_\mu (x) \equiv - \sigma_{\nu \rho} \gamma_\mu {\rm tr}_c ( \,u_{\nu\,\rho}  (x) \lambda (x)), \qquad \sigma_{\nu \rho}=\frac{1}{2} [\gamma_{\nu},\gamma_{\rho}].
\ee

  The supercurrent is the conserved quantity associated with SUSY. When SUSY is broken (as is the case in both dimensional and lattice regularizations), $S_\mu (x)$ is no longer conserved and must be renormalized. $S_\mu (x)$ can mix with other operators of the same or lower dimensionality which respect the same global symmetries. On general grounds~\cite{Joglekar:1975nu}, such operators can be separated into four classes:
  \begin{itemize}
\item {\bf Class G:} Gauge-invariant operators.
\item {\bf Class A:} BRST variations of some operator.
\item {\bf Class B:} Operators which vanish by the equations of motion.
\item {\bf Class C:} Any other operators which share the same global symmetries, but do not belong to the above classes; these can at most have finite mixing with $S_\mu (x)$.
  \end{itemize}
  Mixing with gauge-noninvariant (class A-C) operators results from the introduction of a gauge-fixing term in the action. A list of all possible candidate operators which can mix with $S_\mu (x)$ can be found in Refs.~\cite{Bergner:2021kbg, Bergner:2022wnb}.


\subsection{Supercurrent renormalization in GIRS}

  To simplify the mixing problem, we implement a renormalization scheme in which only gauge-invariant Green's functions are considered; thus a nonperturbative implementation of such a scheme avoids gauge fixing altogether. In particular, by extending the X-space scheme~\cite{Gimenez:2004me, Chetyrkin:2010dx, Cichy:2012is, Tomii:2018zix}, we consider Green's functions of products of gauge-invariant operators (at different spacetime points, in a way as to avoid potential contact singularities), e.g.,
\be
\langle S_\mu (x) \ S_\nu (y) \rangle, \qquad (x \neq y).
\label{GIRScons}
\ee
The case at hand does not require Green's functions containing products of more than two operators, whose evaluation is more demanding, both perturbatively and nonperturbatively. Moreover, the gauge-noninvariant operators of classes A-C cannot contribute to such Green's functions and they need not be considered any further. As a consequence, the set of mixing operators in GIRS is greatly reduced and includes only gauge-invariant operators, which are accessible by lattice simulations.

There is only one gauge-invariant operator $T_\mu (x)$ (see Ref.~\cite{Ali:2018fbq} and references therein), which can mix with the supercurrent:
\be
T_\mu (x) \equiv 2\,\gamma_\nu {\rm tr}_c(\,u_{\mu\,\nu} (x) \lambda (x)). 
\ee
Thus, we construct a $2 \times 2$ mixing matrix, which relates the bare (B) to the renormalized (R) operators:
\begin{gather}
 \begin{pmatrix} {S}^R_\mu (x) & \\[2ex] {T}^R_{\mu} (x)\end{pmatrix}
 =
\begin{pmatrix}
   Z_{SS}^{B,R} &
   Z_{ST}^{B,R} \\[2ex]
   Z_{TS}^{B,R} &
   Z_{TT}^{B,R} 
   \end{pmatrix}
   \begin{pmatrix} {S}^B_{\mu} (x) & \\[2ex] {T}^B_{\mu} (x)\end{pmatrix}.
   \label{mixing matrix}
\end{gather}
[The index $B$  refers to dimensional (${\rm DR}$)  or lattice (${\rm L}$) regularization, and $R = \MSbar, {\rm GIRS}$ to the renormalization scheme. For simplicity, the renormalization functions 
$Z^{B,R}$ will be often denoted merely as $Z$.]

The determination of the four elements of the mixing matrix requires four conditions. A maximum of three conditions can be imposed by considering expectation values between the two mixing operators:
\bea
G^{SS}_{\mu \nu}(x, y) &\equiv& \langle S_\mu (x) \ \overline{ S}_\nu (y) \rangle,
\label{GFSS}
\\
G^{TT}_{\mu \nu}(x, y) &\equiv& \langle T_\mu (x) \ \overline{ T}_\nu (y) \rangle,
\label{GFTT}
\\
G^{ST}_{\mu \nu}(x, y) &\equiv& \langle S_\mu (x) \ \overline{ T}_\nu (y) \rangle,
\label{GFST}
\\
G^{TS}_{\mu \nu}(x, y) &\equiv& \langle T_\mu (x) \ \overline{ S}_\nu (y) \rangle,
\label{GFTS}
\eea
where
\bea
\overline{ S}_\mu (x) &\equiv& \,{\rm tr}_c (\, \bar \lambda (x) u_{\nu\,\rho} (x)) \gamma_\mu \sigma_{\nu \rho},
\\
\overline{ T}_\mu (x) &\equiv& 2\, {\rm tr}_c(\,\bar \lambda (x) u_{\mu\,\nu} (x))\gamma_\nu.
\eea
For convenience, we choose to express the ``bar'' operators in terms of $\bar{\lambda}$ rather than $\lambda$. The bar operators $\overline{A} = \overline{S}_\mu, \overline{T}_\mu$ are related to the original operators $A = S_\mu, T_\mu$ through charge conjugation transformations, as follows:
\be
\overline{A} \equiv A_C^T \ C^T, \qquad A = C \overline{A}_C^T,
\label{baroperators}
\ee
where $A_C$ is the operator $A$ with its fields replaced by their charge conjugates, and $C$ is the charge conjugation matrix satisfying $C \gamma_\mu C^{-1} = - \gamma_\mu^T$. The transformations of fields under charge conjugation are given below:
\bea
u_\mu &\rightarrow& -u_\mu^T, \\
\lambda &\rightarrow& C \overline{\lambda}^T, \\
\overline{\lambda} &\rightarrow& - \lambda^T C^{-1}.
\eea

A fourth condition can be obtained by considering two-point Green's functions involving products of $S_\mu (x)$ or $T_\mu (x)$ with the Gluino-Glue operator $\mathcal{O} (x)$, which is the only other gauge-invariant operator of equal or lower dimension:
\bea
G^{S{\cal O}}_{\mu}(x, y) &\equiv& \langle S_\mu (x) \ \overline{ {\cal O}} (y) \rangle,
\label{GFSGg}
\\
G^{T{\cal O}}_{\mu}(x, y) &\equiv& \langle T_\mu (x) \ \overline{ {\cal O}} (y) \rangle,
\label{GFTGg}
\\
G^{{\cal O}S}_{\mu}(x, y) &\equiv& \langle {\cal O} (x) \ \overline{S}_\mu (y) \rangle,
\label{GFGgS}
\\
G^{{\cal O}T}_{\mu}(x, y) &\equiv& \langle {\cal O} (x) \ \overline{T}_\mu (y) \rangle,
\label{GFGgT}
\eea
where 
\bea
{\cal O} (x) &\equiv& \sigma_{\mu \nu} \,{\rm{tr}}_c (\,u_{\mu \nu} (x) \lambda (x)), \\
\overline{ {\cal O}} (x) &\equiv&  \,{\rm{tr}}_c (\,\bar \lambda (x) u_{\mu \nu} (x)) \sigma_{\mu \nu}.
\label{gluino-glue}
\eea
$\overline{ {\cal O}} (x)$ satisfies Eq. \eqref{baroperators}. The operator ${\cal O} (x)$ is multiplicatively renormalizable in GIRS~\cite{Costa:2021pfu} and its renormalization factor is obtained by considering the following Green's function:
\be
G^{{\cal O}{\cal O}}(x, y) \equiv \langle {\cal O} (x) \ \overline{{\cal O}} (y) \rangle.
\label{GFGgGg}
\ee

Given Eq.~\eqref{baroperators}, the above Green's functions are related among themselves through charge conjugation transformations, as follows:
\be
\left\langle A (x) {\overline{B}} (y) \right\rangle = C {\left\langle B (y) {\overline{A}} (x) \right\rangle}^T C^{-1}, \label{GFsidentity}
\ee
where $A, B = S_\mu, T_\mu, {\cal O}$. Note that for determining the mixing matrix, we may not need to implement all choices of Lorentz/Dirac indices in a nonperturbative evaluation of the above Green's functions; the exact choice of the renormalization conditions will dictate which components of the Green's functions are needed. However, in order to determine a consistent and solvable set of nonperturbative renormalization conditions, we need to calculate all Green's functions perturbatively in dimensional regularization. Also, since there is no unique way of selecting solvable conditions in GIRS, a perturbative calculation of all the above Green's functions will be useful for determining conversion factors from all possible variants of GIRS to the $\MSbar$ scheme.  

\section{Perturbative calculation in dimensional regularization}
\label{Perturbative calculation}

In this Section we investigate the mixing problem in the continuum by regularizing the theory in $d\equiv4-2\varepsilon$ dimensions. We apply both $\MSbar$ and $\GIRS$ and we extract the conversion matrix, $C^{\GIRS,\MSbar}$, between the two schemes to one loop. The conversion matrix along with the lattice mixing matrix in $\GIRS$, $Z^{\LR,\GIRS}$, (computed nonperturbatively) allow us to determine the lattice mixing matrix in the $\MSbar$ scheme, $Z^{\LR,\MSbar}$, through the following relation:

\begin{gather}
 \begin{pmatrix}
   Z_{SS}^{\LR,\MSbar} &
   Z_{ST}^{\LR,\MSbar} \\[2ex]
   Z_{TS}^{\LR,\MSbar} &
   Z_{TT}^{\LR,\MSbar} 
 \end{pmatrix}
 =
  \begin{pmatrix}
   C_{SS}^{\GIRS,\MSbar} &
   C_{ST}^{\GIRS,\MSbar} \\[2ex]
   C_{TS}^{\GIRS,\MSbar} &
   C_{TT}^{\GIRS,\MSbar} 
  \end{pmatrix}
  \begin{pmatrix}
   Z_{SS}^{\LR,\GIRS} &
   Z_{ST}^{\LR,\GIRS} \\[2ex]
   Z_{TS}^{\LR,\GIRS} &
   Z_{TT}^{\LR,\GIRS} 
 \end{pmatrix}.
   \label{conversion matrix}
\end{gather}

\subsection{One-loop calculation in the \texorpdfstring{$\MSbar$}{MS bar} scheme} 

\begin{figure}[ht]
\centering 
\includegraphics[width=.09\textwidth]{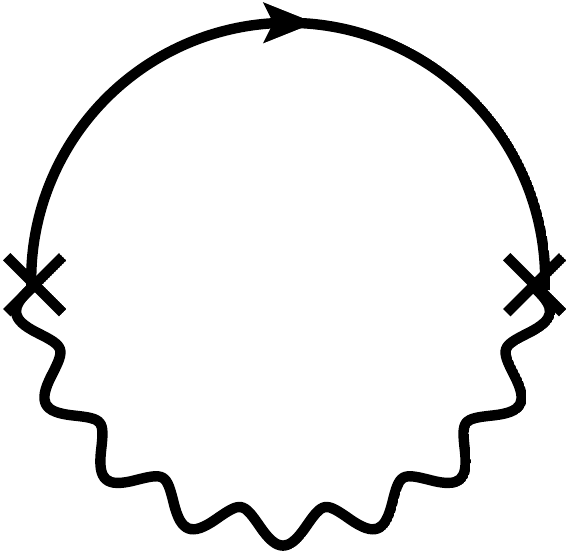}
\hfil
\includegraphics[width=.45\textwidth]{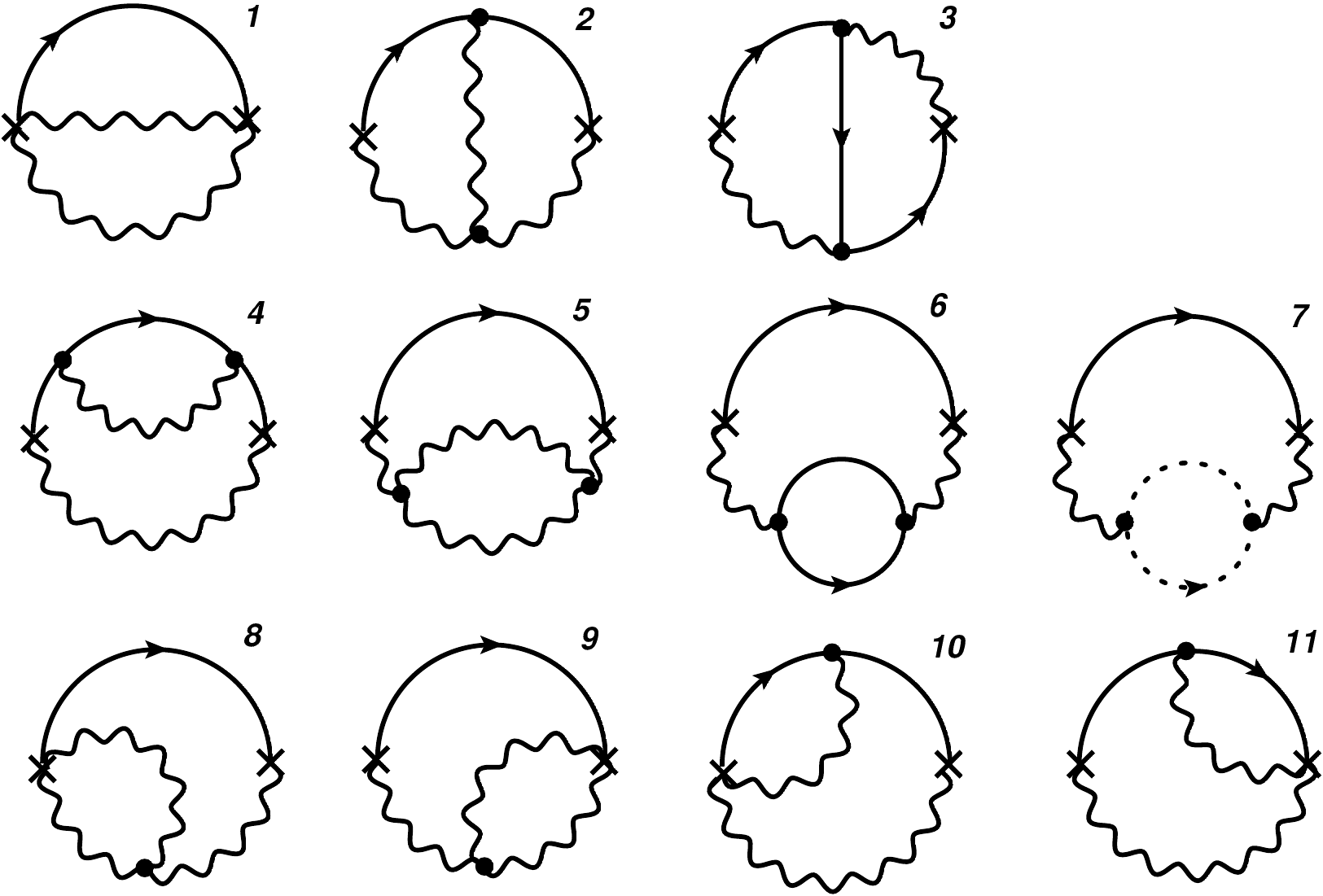}
\caption{One-loop and two-loop Feynman diagrams contributing to the tree-level and one-loop two-point Green's functions of Eqs. (\ref{GFSS} -- \ref{GFTS}, \ref{GFSGg} -- \ref{GFGgT}, \ref{GFGgGg}). A wavy (solid, dashed) line represents gluons (gluinos, ghosts). The two crosses denote the insertions of operators $S_\mu,T_\nu, \mathcal{O}$ appearing in the definition of each two-point function.}
\label{diagrams}
\end{figure}

First, we present our results for the $\MSbar$-renormalized Green's functions. Due to the $2\times2$ mixing, the renormalized Green's functions are linear combinations of bare Green's functions, which are given below in matrix form:
\bea
  \begin{pmatrix}
    G^{SS, \, R}_{\mu\nu} &
    G^{ST, \,R}_{\mu\nu} \\[2ex]
    G^{TS, \,R}_{\mu\nu} &
    G^{TT, \,R}_{\mu\nu}
  \end{pmatrix}
  &=& \quad \ \ \
   \begin{pmatrix}
   Z_{SS} &
   Z_{ST} \\[2ex]
   Z_{TS} &
   Z_{TT}
   \end{pmatrix}
   \begin{pmatrix}
    G^{SS, \,B}_{\mu\nu} &
    G^{ST, \,B}_{\mu\nu} \\[2ex]
    G^{TS, \,B}_{\mu\nu} &
    G^{TT, \,B}_{\mu\nu}
   \end{pmatrix}
   \begin{pmatrix}
   Z_{SS} &
   Z_{TS} \\[2ex]
   Z_{ST} &
   Z_{TT}
   \end{pmatrix}, \\
   \nonumber \\
   \begin{pmatrix}
     G^{S\mathcal{O}, \,R}_{\mu} \\[2ex]
     G^{T\mathcal{O}, \,R}_{\mu}
   \end{pmatrix}
   &=&
   Z_{\mathcal{O}} \
   \begin{pmatrix}
   Z_{SS} &
   Z_{ST} \\[2ex]
   Z_{TS} &
   Z_{TT}
   \end{pmatrix}
   \begin{pmatrix}
     G^{S\mathcal{O}, \,B}_{\mu} \\[2ex]
     G^{T\mathcal{O}, \,B}_{\mu}
   \end{pmatrix}, \\
      \nonumber \\
   \begin{pmatrix}
     G^{\mathcal{O}S, \,R}_{\mu} \\[2ex]
     G^{\mathcal{O}T, \,R}_{\mu}
   \end{pmatrix}
   &=&
    Z_{\mathcal{O}} \
   \begin{pmatrix}
   Z_{SS} &
   Z_{ST} \\[2ex]
   Z_{TS} &
   Z_{TT}
   \end{pmatrix}
   \begin{pmatrix}
     G^{\mathcal{O}S, \,B}_{\mu} \\[2ex]
     G^{\mathcal{O}T, \,B}_{\mu}
   \end{pmatrix}, \\
      \nonumber \\
   G^{\mathcal{O}\mathcal{O}, \,R} &=& (Z_{\mathcal{O}})^2 \ \ G^{\mathcal{O}\mathcal{O}, \,B}.
   \eea
   The evaluation of the corresponding bare Green's functions at tree-level and to one-loop order in dimensional regularization (DR) involve, respectively, the one-loop and two-loop Feynman diagrams of Fig.~\ref{diagrams}.

As is standard practice, the pole terms ($1/\varepsilon^n$, $n \in \mathbb{Z}^+$) are removed by defining the $\MSbar$ mixing matrix elements to have only negative integer powers of $\varepsilon$, i.e., $Z_{ij}^{\DR,\MSbar} = \delta_{ij} + g^2 (z_{ij} / \varepsilon) + \mathcal{O} (g^4)$, where $i, j = S, T$ and $Z_{\mathcal{O}}^{\DR,\MSbar} = 1 + g^2 (z_{\mathcal{O}} / \varepsilon) + \mathcal{O} (g^4)$. Our results for $Z_{ij}^{\DR,\MSbar}$, $Z_{\mathcal{O}}^{\DR,\MSbar}$ read:
\bea
Z_{SS}^{\DR,\MSbar} &=& 1 + \mathcal{O} (g^4), \label{ZSSMSbar} \\
Z_{ST}^{\DR,\MSbar} &=& \mathcal{O} (g^4), \\
Z_{TS}^{\DR,\MSbar} &=& \gtilde \frac{3 N_c}{2 \varepsilon} + \mathcal{O} (g^4), \\
Z_{TT}^{\DR,\MSbar} &=& 1 - \gtilde \frac{3 N_c}{\varepsilon} + \mathcal{O} (g^4), \label{ZTTMSbar} \\
Z_{\mathcal{O}}^{\DR,\MSbar} &=& 1 - \gtilde \frac{3 N_c}{\varepsilon} + \mathcal{O} (g^4),
\eea
which agree with our recent one-loop calculations of Refs.~\cite{Costa:2020keq, Bergner:2022wnb}, where gauge-noninvariant Green's functions of all mixing operators with external elementary fields are considered. The one-loop expressions for the $\MSbar$-renormalized Green's functions are given below:
\bea
G^{SS, \,\MSbar}_{\mu \nu}(x, y) &=& -\frac{2 (N_c^2 -1)}{3 \pi^4 {(z^2)}^4} \, (3 - 5 \frac{g^2_{\MSbar}}{16 \pi^2} N_c) \, (4 s_{\mu \nu}^{[3]} + s_{\mu \nu}^{[4]}), \label{GSSMSbar}\\
G^{TT, \,\MSbar}_{\mu \nu}(x, y) &=& -\frac{(N_c^2 -1)}{6 \pi^4 {(z^2)}^4} \Big[(3 - 5 \frac{g^2_{\MSbar}}{16 \pi^2} N_c) \, (4 s_{\mu \nu}^{[3]} - 2 s_{\mu \nu}^{[4]}) \nonumber \\
&& \qquad \qquad \qquad + 9 \frac{g^2_{\MSbar}}{16 \pi^2} N_c \Big(2 s_{\mu \nu}^{[1]} +2 s_{\mu \nu}^{[2]} - 3\, (3 + 4 \gamma_E + 2 \ln (\bar{\mu}^2 z^2/4)) \, s_{\mu \nu}^{[4]}\Big)  \Big], \\
G^{ST, \,\MSbar}_{\mu \nu}(x, y) &=& -\frac{(N_c^2 -1)}{3 \pi^4 {(z^2)}^4} \Big[(3 - 5 \frac{g^2_{\MSbar}}{16 \pi^2} N_c) (4 s_{\mu \nu}^{[3]} + s_{\mu \nu}^{[4]}) + 18 \frac{g^2_{\MSbar}}{16 \pi^2} N_c (s_{\mu \nu}^{[2]} - s_{\mu \nu}^{[4]})\Big], \\
G^{TS, \,\MSbar}_{\mu \nu}(x, y) &=& -\frac{(N_c^2 -1)}{3 \pi^4 {(z^2)}^4} \Big[(3 - 5 \frac{g^2_{\MSbar}}{16 \pi^2} N_c) (4 s_{\mu \nu}^{[3]} + s_{\mu \nu}^{[4]}) + 18 \frac{g^2_{\MSbar}}{16 \pi^2} N_c (s_{\mu \nu}^{[1]} - s_{\mu \nu}^{[4]})\Big], \\
G^{S{\cal O}, \,\MSbar}_{\mu}(x, y) &=& -\frac{(N_c^2 -1)}{\pi^4 {(z^2)}^4} 12 \frac{g^2_{\MSbar}}{16 \pi^2} N_c s_{\mu}^{[6]}, \\
G^{{\cal O}S, \,\MSbar}_{\mu}(x, y) &=& -\frac{(N_c^2 -1)}{\pi^4 {(z^2)}^4} 12 \frac{g^2_{\MSbar}}{16 \pi^2} N_c s_{\mu}^{[6]}, \\
G^{T{\cal O}, \,\MSbar}_{\mu}(x, y) &=& -\frac{(N_c^2 -1)}{\pi^4 {(z^2)}^4} \Big[\Big(3 + 2 \frac{g^2_{\MSbar}}{16 \pi^2} N_c (8 + 18 \gamma_E + 9 \ln (\bar{\mu}^2 z^2/4))\Big) (\phantom{+}s_{\mu}^{[5]} + s_{\mu}^{[6]}) - 6 \frac{g^2_{\MSbar}}{16 \pi^2} N_c s_{\mu}^{[5]} \Big], \\
G^{{\cal O}T, \,\MSbar}_{\mu}(x, y) &=& -\frac{(N_c^2 -1)}{\pi^4 {(z^2)}^4} \Big[\Big(3 + 2 \frac{g^2_{\MSbar}}{16 \pi^2} N_c (8 + 18 \gamma_E + 9 \ln (\bar{\mu}^2 z^2/4))\Big) (-s_{\mu}^{[5]} + s_{\mu}^{[6]}) + 6 \frac{g^2_{\MSbar}}{16 \pi^2} N_c s_{\mu}^{[5]} \Big], \\
G^{{\cal O}{\cal O}, \,\MSbar}(x, y) &=& -\frac{2 (N_c^2 -1)}{\pi^4 {(z^2)}^4} \Big[3 + 2 \frac{g^2_{\MSbar}}{16 \pi^2} N_c (5 + 18 \gamma_E + 9 \ln (\bar{\mu}^2 z^2/4))\Big] \slashed{z}\label{GOOMSbar},
\eea
where $z \equiv x - y$,
  \bea
  s_{\mu \nu}^{[1]} (z) &\equiv& \gamma_\mu z_\nu, \qquad s_{\mu \nu}^{[2]} (z) \equiv \gamma_\nu z_\mu, \qquad s_{\mu \nu}^{[3]} (z) \equiv (\delta_{\mu \nu} - 2 \frac{z_\mu z_\nu}{z^2}) \slashed{z}, \label{struct1}\\
  s_{\mu \nu}^{[4]} (z) &\equiv& \gamma_\mu \slashed{z} \gamma_\nu, \quad \ \, s_{\mu}^{[5]} (z) \equiv z_\mu \mathbf{1}, \qquad \ \, s_{\mu}^{[6]} (z) \equiv \sigma_{\mu \rho} z_\rho. \label{struct2}
  \eea
  If a gluino mass were present the above Green's functions would also contain the structures of Eqs. (\ref{struct1} -- \ref{struct2}) multiplied by an extra $\slashed{z}$. 

\subsection{Renormalization conditions in GIRS}
  
  Next, we define appropriate renormalization conditions in GIRS, which must be applicable in both continuum and lattice. There is, {\it a priori}, wide flexibility in determining conditions in GIRS. In our study we consider the following set of conditions, in which we integrate over the spatial components of $z = x - y = (\vec{z},t)$\,:  
\bea
\int d^3 \vec{z} \ {\rm Tr} \left[G^{SS, \,\GIRS}_{\mu \nu}(x, y) \ P_{\nu \mu}\right] &=& \int d^3 \vec{z} \ {\rm Tr} \left[G^{SS, \,{\rm tree}}_{\mu \nu}(x, y) \ P_{\nu \mu} \right], \label{GIRS2_cond1} \\
\int d^3 \vec{z} \ {\rm Tr} \left[G^{TT, \,\GIRS}_{\mu \nu}(x, y) \ P_{\nu \mu} \right] &=& \int d^3 \vec{z} \ {\rm Tr} \left[G^{TT, \,{\rm tree}}_{\mu \nu}(x, y) \ P_{\nu \mu} \right], \label{GIRS2_cond2} \\
\int d^3 \vec{z} \ {\rm Tr} \left[G^{ST, \,\GIRS}_{\mu \nu}(x, y) \ P_{\nu \mu} \right] &=& \int d^3 \vec{z} \ {\rm Tr} \left[G^{ST, \,{\rm tree}}_{\mu \nu}(x, y) \ P_{\nu \mu} \right], \label{GIRS2_cond3} \\
\int d^3 \vec{z} \ {\rm Tr} \left[G^{S{\cal O}, \,\GIRS}_{\mu}(x, y) \ P_{\mu} \ \right] &=& \int d^3 \vec{z} \ {\rm Tr} \left[G^{S{\cal O}, \,{\rm tree}}_{\mu}(x, y) \ P_{\mu} \ \right].
\label{GIRS2_cond4}
\eea
where
\be
P_{\nu \mu} = \gamma_\nu \gamma_4 \gamma_\mu, \qquad P_{\mu} = \gamma_4 \gamma_\mu,
\label{proj}
\ee
and the repeated indices $\mu, \nu$ are not summed over. The tree-level values in the r.h.s. of Eqs. (\ref{GIRS2_cond1} -- \ref{GIRS2_cond4}) are given below:
  \bea
  \int d^3 \vec{z} \ {\rm Tr} \left[G^{SS, \,{\rm tree}}_{\mu \nu}(x, y) \ P_{\nu \mu} \right] &=& - \frac{(N_c^2 - 1) \ t}{\pi^2 |t|^5} (1 - \delta_{\mu 4} - \delta_{\nu 4} - 3 \,\delta_{\mu \nu} + 4\, \delta_{\mu 4} \,\delta_{\nu 4}), \label{TL1GIRS2} \\
  \int d^3 \vec{z} \ {\rm Tr} \left[G^{TT, \,{\rm tree}}_{\mu \nu}(x, y) \ P_{\nu \mu} \right] &=& \phantom{+} \frac{(N_c^2 - 1) \ t}{4 \pi^2 |t|^5} (2 + \delta_{\mu 4} + \delta_{\nu 4}  + 3 \,\delta_{\mu \nu} - 4 \,\delta_{\mu 4}\,\delta_{\nu 4}), \\
  \int d^3 \vec{z} \ {\rm Tr} \left[G^{ST, \,{\rm tree}}_{\mu \nu}(x, y) \ P_{\nu \mu} \right] &=& - \frac{(N_c^2 - 1) \ t}{2 \pi^2 |t|^5} (1 - \delta_{\mu 4} - \delta_{\nu 4} - 3 \,\delta_{\mu \nu} + 4\, \delta_{\mu 4}\,\delta_{\nu 4}), \label{TL3GIRS2}\\
  \int d^3 \vec{z} \ {\rm Tr} \left[G^{S\mathcal{O}, \,{\rm tree}}_{\mu}(x, y) \ P_{\mu} \right] \ &=& \phantom{+} 0.
  \eea

On the lattice, integration over timeslices is replaced by summation, which is expected to reduce statistical errors in the nonpertubative data of the numerical simulations. The explicit form of the corresponding conditions on the lattice can be found in the next section.

Alternative definitions of the GIRS conditions may involve higher moments of the Green's functions, e.g., $\int d^3 \vec{z}  \ z_\mu z_\nu {\rm Tr} \left[G^{A\,B}_{\mu \nu}  P_{\nu \mu} \right], (A,B = S, T)$. Such choices can affect statistical error either positively, since some contributions from positive and negative directions will now add up, rather than cancel, or negatively, since contributions from larger values of $z$, which are expected to be more noisy, will now appear multiplied by extra powers of $z$.

While in $\MSbar$ the renormalization factors for each operator are independent of Lorentz components, in GIRS different choices of the Lorentz components (spatial or temporal) for each operator can, {\it in principle}, give different renormalization factors. Thus, in order to determine a consistent set of conditions in GIRS, we need to use the same components (spatial or temporal) for each operator in all conditions. Also, due to the integration over the spatial components of $z$, the possible choices of Lorentz indices in each operator which give a solution to the system of conditions are further limited. By writing down the conditions of Eqs. (\ref{GIRS2_cond1} -- \ref{GIRS2_cond4}) in terms of bare Green's functions\footnote{Integration over spatial components of $z$ is understood in each trace.}:
  \bea
&& Z_{SS}^2 \ {\rm Tr} \left[G^{SS}_{\mu \nu} P_{\nu \mu} \right] + Z_{SS} \ Z_{ST} \ ({\rm Tr} \left[G^{ST}_{\mu \nu} P_{\nu \mu} \right] + {\rm Tr} \left[G^{TS}_{\mu \nu} P_{\nu \mu} \right]) + Z_{ST}^2 \ {\rm Tr} \left[G^{TT}_{\mu \nu} P_{\nu \mu} \right] = {\rm Tr} \left[G^{SS,{\rm tree}}_{\mu \nu} P_{\nu \mu} \right], \label{cond1_explicit} \\
&& Z_{TS}^2 \ {\rm Tr} \left[G^{SS}_{\mu \nu} P_{\nu \mu} \right] + Z_{TS} \ Z_{TT} \ ({\rm Tr} \left[G^{ST}_{\mu \nu} P_{\nu \mu} \right] + {\rm Tr} \left[G^{TS}_{\mu \nu} P_{\nu \mu} \right]) + Z_{TT}^2 \ {\rm Tr} \left[G^{TT}_{\mu \nu} P_{\nu \mu} \right] = {\rm Tr} \left[G^{TT,{\rm tree}}_{\mu \nu} P_{\nu \mu} \right], \label{cond2_explicit} \\
  && Z_{SS} \ \left(Z_{TS} \ {\rm Tr} \left[G^{SS}_{\mu \nu} P_{\nu \mu} \right] + Z_{TT} \ {\rm Tr} \left[G^{ST}_{\mu \nu} P_{\nu \mu} \right]\right) + Z_{ST} \ \left(Z_{TS} \ {\rm Tr} \left[G^{TS}_{\mu \nu} P_{\nu \mu} \right] + Z_{TT} \ {\rm Tr} \left[G^{TT}_{\mu \nu} P_{\nu \mu} \right]\right) = \nonumber \\
 && \qquad \qquad \qquad \qquad \qquad \qquad \qquad \qquad \qquad \qquad \qquad \qquad \qquad \qquad \qquad \qquad \qquad \qquad \qquad {\rm Tr} \left[G^{ST,{\rm tree}}_{\mu \nu} P_{\nu \mu} \right], \label{cond3_explicit} \\
&& Z_O \left(Z_{SS} \ {\rm Tr} \left[G^{SO}_{\mu} P_{\mu} \right] + Z_{ST} \ {\rm Tr} \left[G^{TO}_{\mu} P_{\mu} \right]\right) = {\rm Tr} \left[G^{SO,{\rm tree}}_{\mu} P_{\mu} \right] = 0, \label{cond4_explicit}
 \eea
 we conclude that indices $\mu$ and $\nu$ must be both spatial; the choice of a temporal component in Eq. \eqref{cond1_explicit} gives a vanishing tree-level value (see Eq. \eqref{TL1GIRS2}), which in combination with Eq. \eqref{cond4_explicit} results in either vanishing or indeterminate values for $Z_{SS}$ and $Z_{ST}$. There is no restriction in choosing which spatial components $\mu$ and $\nu$ will be in each condition (same or different). However, some choices may be preferable from the simulation point of view. In particular, we set: $\mu = \nu = i$ in Eq. \eqref{cond1_explicit}, $\mu = \nu = j$ in Eq. \eqref{cond2_explicit}, $\mu = \nu = k$ in Eq. \eqref{cond3_explicit} and $\mu = \ell$ in Eq. \eqref{cond4_explicit}, where $i,j,k, \ell$ can be equal or different among themselves (of course, at least two of the four indices will be equal). These choices of indices give larger signal-to-noise ratios, as is observed in our nonperturbative study described in the next section.  Also, averages over $i,j,k, \ell$ can be employed for improving the signal.


Eq. \eqref{cond1_explicit} and Eq. \eqref{cond4_explicit} lead to a second degree equation from which $Z_{SS}$ (and subsequently $Z_{ST}$) can be determined. Following that, Eq. \eqref{cond2_explicit} and Eq. \eqref{cond3_explicit} lead to another 
second degree equation for $Z_{TT}$ (and also $Z_{TS}$). We thus obtain:
  \bea
 Z_{SS} &=& \sqrt{2 c \ {\{ {\rm Tr} \left[G^{SS}_{ii} P_{ii} \right] - ({\rm Tr} \left[G^{ST}_{ii} P_{ii} \right] + {\rm Tr} \left[G^{TS}_{ii} P_{ii} \right]) R_1 + {\rm Tr} \left[G^{TT}_{ii} P_{ii} \right] R_1^2 \}}^{-1}}, \label{sol1} \\
 Z_{ST} &=& - Z_{SS} R_1, \\
 Z_{TT} &=& R_4 + \sqrt{R_4^2 + R_5}, \\ 
 Z_{TS} &=& Z_{SS}^{-1} R_2 - Z_{TT} R_3, \label{sol4}
  \eea
  where
  \bea
  c &\equiv& \frac{(N_c^2 - 1) t}{\pi^2 |t|^5}, \\
  \nonumber \\
  R_1 &\equiv& \frac{{\rm Tr} \left[G^{SO}_{\ell} P_{\ell} \right]}{{\rm Tr} \left[G^{TO}_{\ell} P_{\ell} \right]}, \label{R1} \\
  \nonumber \\
  R_2 &\equiv& \frac{c}{{\rm Tr} \left[G^{SS}_{kk} P_{kk} \right] - {\rm Tr} \left[G^{TS}_{kk} P_{kk} \right] R_1}, \label{R2} \\
    \nonumber \\
    R_3 &\equiv& \frac{{\rm Tr} \left[G^{ST}_{kk} P_{kk} \right] - {\rm Tr} \left[G^{TT}_{kk} P_{kk} \right] R_1}{{\rm Tr} \left[G^{SS}_{kk} P_{kk} \right] - {\rm Tr} \left[G^{TS}_{kk} P_{kk} \right] R_1}, \label{R3} \\
      \nonumber \\
      R_4 &\equiv& \frac{R_2 \ \{ 2 {\rm Tr} \left[G^{SS}_{jj} P_{jj} \right] R_3 - ({\rm Tr} \left[G^{ST}_{jj} P_{jj} \right] + {\rm Tr} \left[G^{TS}_{jj} P_{jj} \right]) \}}{2 Z_{SS} \ \{ {\rm Tr} \left[G^{SS}_{jj} P_{jj} \right] R_3^2 - ({\rm Tr} \left[G^{ST}_{jj} P_{jj} \right] + {\rm Tr} \left[G^{TS}_{jj} P_{jj} \right]) R_3 + {\rm Tr} \left[G^{TT}_{jj} P_{jj} \right] \}}, \label{R4} \\
        \nonumber \\
  R_5 &\equiv& \frac{(5 / 4) \ c - {\rm Tr} \left[G^{SS}_{jj} P_{jj} \right] Z_{SS}^{-2} \ R_2^2}{{\rm Tr} \left[G^{SS}_{jj} P_{jj} \right] R_3^2 - ({\rm Tr} \left[G^{ST}_{jj} P_{jj} \right] + {\rm Tr} \left[G^{TS}_{jj} P_{jj} \right]) R_3 + {\rm Tr} \left[G^{TT}_{jj} P_{jj} \right]}. \label{R5} 
  \eea
  The choice of plus/minus signs for the square roots appearing in the solutions of the second degree equations (Eqs. (\ref{cond1_explicit}--\ref{cond4_explicit})) is dictated by the requirement that the Z matrix be equal to the unit matrix at tree level. Because of the zero tree-level value in the r.h.s. of Eq. \eqref{cond4_explicit}, the renormalization factor $Z_O$ is eliminated and thus does not enter the solution.

\subsection{Results for the conversion matrix}

The final step in our perturbative calculation is the evaluation of the conversion matrix $C^{\GIRS, \MSbar}$ between $\GIRS$ and $\MSbar$ scheme. Given that the conversion factors between two renormalization schemes are regularization-independent, we calculate $C^{\GIRS, \MSbar}$ by using results for the mixing matrix in dimensional regularization, instead of lattice regularization, through the following relation:
\bea
\begin{pmatrix}
C_{SS}^{{\rm GIRS},\overline{\rm MS}} & C_{ST}^{{\rm GIRS},\overline{\rm MS}} \\
\\
C_{TS}^{{\rm GIRS},\overline{\rm MS}} & C_{TT}^{{\rm GIRS},\overline{\rm MS}}
\end{pmatrix} &=& \begin{pmatrix}
Z_{SS}^{{\rm L},\overline{\rm MS}} & Z_{ST}^{{\rm L},\overline{\rm MS}} \\
\\
Z_{TS}^{{\rm L},\overline{\rm MS}} & Z_{TT}^{{\rm L},\overline{\rm MS}}
\end{pmatrix} \cdot \begin{pmatrix}
Z_{SS}^{{\rm L},{\rm GIRS}} & Z_{ST}^{{\rm L},{\rm GIRS}} \\
\\
Z_{TS}^{{\rm L},{\rm GIRS}} & Z_{TT}^{{\rm L},{\rm GIRS}}
\end{pmatrix}^{-1} \nonumber \\
&=& \begin{pmatrix}
Z_{SS}^{{\rm DR},\overline{\rm MS}} & Z_{ST}^{{\rm DR},\overline{\rm MS}} \\
\\
Z_{TS}^{{\rm DR},\overline{\rm MS}} & Z_{TT}^{{\rm DR},\overline{\rm MS}}
\end{pmatrix} \cdot \begin{pmatrix}
Z_{SS}^{{\rm DR},{\rm GIRS}} & Z_{ST}^{{\rm DR},{\rm GIRS}} \\
\\
Z_{TS}^{{\rm DR},{\rm GIRS}} & Z_{TT}^{{\rm DR},{\rm GIRS}}
\end{pmatrix}^{-1}.
\eea
By combining our one-loop results for the mixing matrix in $\GIRS$ (Eqs. (\ref{sol1} -- \ref{sol4})) and $\MSbar$ (Eqs. (\ref{ZSSMSbar} -- \ref{ZTTMSbar})), we  extract the one-loop conversion matrix elements \footnote{Terms of the form $\ln (\bar{\mu}^2 t^2)$ have arisen as a result of integrating $\ln (\bar{\mu}^2 z^2 / 4) / (z^2)^4$ (cf. Eqs. (\ref{GSSMSbar}--\ref{GOOMSbar})) over spatial z-components.}: 

  \bea
    C_{SS}^{{\rm GIRS},\overline{\rm MS}} &=& 1 - \frac{g^2_{\MSbar}}{16 \pi^2} \frac{17 N_c}{6} + \mathcal{O} (g^4_{\MSbar}), \label{convSS2}\\
    C_{ST}^{{\rm GIRS},\overline{\rm MS}} &=& \frac{g^2_{\MSbar}}{16 \pi^2} 4 N_c + \mathcal{O} (g^4_{\MSbar}), \\
    C_{TS}^{{\rm GIRS},\overline{\rm MS}} &=& -\frac{g^2_{\MSbar}}{16 \pi^2} \frac{3 N_c }{2} \left(\frac{2}{3} + 2 \gamma_E + \ln (\bar{\mu}^2 t^2)\right) + \mathcal{O} (g^4_{\MSbar}), \\
    C_{TT}^{{\rm GIRS},\overline{\rm MS}} &=& 1 + \frac{g^2_{\MSbar}}{16 \pi^2} N_c \left(\frac{7}{6} + 6 \gamma_E + 3 \ln (\bar{\mu}^2 t^2)\right) + \mathcal{O} (g^4_{\MSbar}). \label{convTT2}
  \eea
  The results in Eqs. (\ref{convSS2}--\ref{convTT2}) are independent of the choices of spatial indices ($i,j,k,\ell$) in the GIRS renormalization conditions.  

\section{Nonperturbative calculation}
\label{Nonperturbative calculation}

The GIRS renormalization conditions  (Eqs. (\ref{cond1_explicit}--\ref{cond4_explicit})), involving only Green's functions between gauge-invariant operators, provide a scheme that can be used for a nonperturbative determination of the supercurrent renormalization factors, Eq. \eqref{mixing matrix}, on the lattice. The conversion matrix (Eqs. (\ref{convSS2}--\ref{convTT2})) allows to translate the results from  GIRS to the $\MSbar$ scheme.

In the following we apply this approach on a first set of test ensembles selected from the ones presented in \cite{Bergner:2015adz,Ali:2019agk}. For comparison, we consider the two gauge groups SU($2$) and SU($3$) and two different lattice actions.
The simulations of SU($2$) SYM have been done with a tree level Symanzik improved gauge action and unimproved Wilson fermions ($c_{\rm SW}=0$, $r=1$). One level of stout smearing is used for the gauge links in the Dirac-Wilson operator in the SU(2) case. The SU($3$) simulations use an unimproved gauge action and clover improved Wilson fermions where $c_{\rm SW}$ determined by one-loop perturbation theory \cite{Musberg:2013foa}.
The lattice action has been introduced in Eq. \eqref{susylagrLattice}.
The lattice Wilson operator is represented in terms of the hopping parameter $\kappa\equiv 1/(2m_0+8)$
\begin{equation}
    D_W = 1 - \kappa \big[(1-\gamma_\mu)(V_\mu(x))\delta_{x+\mu,y} + (1+\gamma_\mu)(V^\dagger{}_\mu(x-\mu))\delta_{x-\mu,y}\big],
    \label{eq:Dirac_Wilson}
\end{equation}
where the links $V_\mu(x)$ in the adjoint representation are given by $V_\mu^{ab}=2\,{\rm tr_c}[U^{\dagger}_\mu (x)T^a U_\mu(x)T^b]$. The clover term is included in a similar way. 

Details of the ensembles of gauge configurations can be found in earlier publications \cite{Bergner:2013nwa,Bergner:2015adz}. For SU($2$) SYM we selected ensembles of two different lattice sizes, one with $V\equiv L^3\times T =24^3 \times 48$ and a larger one with $V=32^3\times64$. Anti-periodic boundary conditions have been implemented in the temporal direction and the gauge coupling constant $\beta=2N_c/g^2$ is $\beta=1.75$ in both cases. In order to investigate possible effects of the finite gluino mass, we considered $\kappa=0.14920$ and $\kappa=0.14925$ for the small lattice, while for the larger one we have selected $\kappa=0.1494$. In case of SU($3$) SYM we selected only a single test ensemble with $\beta=5.6$,  $\kappa=0.1655$, $c_{\rm SW}=1.587$, and $V=24^3 \times 48$.

The Dirac-Wilson operator breaks supersymmetry and chiral symmetry, but the critical point $\kappa_c(\beta)$ can be extrapolated according to signals for a vanishing renormalized gluino mass. Supersymmetry and chiral symmetries are recovered in the continuum limit at this point. As explained in our previous studies, the adjoint pion mass provides a signal for chiral symmetry breaking. In the selected ensembles, the pion mass and lattice artefacts are already quite small and nearly degenerate supersymmetry multiplets are observed. For reference, we provide the parameters in Table~\ref{tab:renormalization_factors}.

The parameters of these ensembles have been checked extensively in previous investigations. Finite size effects and the Pfaffian sign are under control. The lattice spacing is small enough to induce only a rather small supersymmetry breaking, while effects like topological freezing become relevant at even smaller lattice spacings \cite{Bergner:2015adz}.

\subsection{Correlators}
The supercurrent operators $S_\mu, T_\mu$ and the gluino-glue operator $\cal{O}$ are generated from clover plaquettes $\hat{F}_{\mu\nu}^{\alpha\beta}(x,t)$ and gluino fields. Their correlators, after integrating out the gluino fields and omitting Lorentz and color indices for clarity, take the following form
\begin{equation}
   \langle A(t)\overline{B}(0)\rangle \equiv \sum_{\vec{x},\vec{y}}\langle A(\vec{x},t)\overline{B}(\vec{y},0)\rangle = \sum_{\vec{x},\vec{y}} \langle\Tr[\Gamma \hat{F}(\vec{x},t)D^ {-1}(\vec{x},t|\vec{y},0)\hat{F}(\vec{y},0)\Gamma']\rangle,
\end{equation}
where $\Gamma, \Gamma'$ are combinations of gamma matrices coming from the respective operators $A, B = S_\mu, T_\mu, {\cal O}$ and the inverse of the Dirac operator $D^{-1}(\vec{x},t|\vec{y},0)$ propagates a gluino from the point $x$ to $y$ 
.To estimate these correlators, we introduced wall sources while the sinks were summed over the spatial positions. 

It is worth noting that the definition of the gluino-glue operator, Eq. \eqref{gluino-glue}, includes both spatial and temporal directions for the gauge links. This symmetric choice of the operator $\mathcal{O}$ is in contrast to earlier works; in particular it is different from the insertion operator chosen in the WI analysis \cite{farchioni_supersymmetric_2002, Ali:2018fbq}.  The symmetric choice is, however, more natural in GIRS as only operators that transform as fully covariant tensors mix with the supercurrent.

\begin{figure}
\begin{minipage}{0.5\textwidth}
\begin{tikzpicture}
  \node (img)  {\includegraphics[width=.75\textwidth]{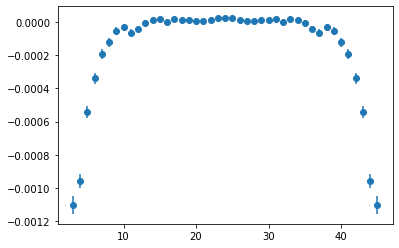}};
  \node[below=of img, node distance=0cm, xshift=0.5cm, yshift=1cm] {t};
  \node[left=of img, node distance=0cm, rotate=90, anchor=center,yshift=-0.7cm,] {$\text{Tr}\langle\mathcal{O}(t)S(0)P\rangle$};
 \end{tikzpicture}
\end{minipage}%
\begin{minipage}{0.5\textwidth}
\begin{tikzpicture}
  \node (img)  {\includegraphics[width=.75\textwidth]{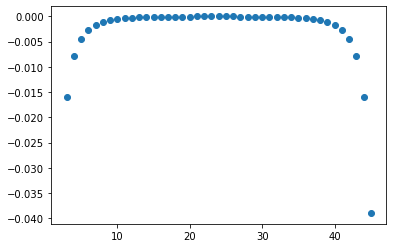}};
  \node[below=of img, node distance=0cm, xshift=0.5cm, yshift=1cm] {t};
  \node[left=of img, node distance=0cm, rotate=90, anchor=center,yshift=-0.7cm,] {$\text{Tr}\langle\mathcal{O}(t)T(0)P\rangle$};
 \end{tikzpicture}
\end{minipage}%
\caption{Correlators $\text{Tr}\langle\mathcal{O}(t)S_i(0)P_i\rangle$ and $\text{Tr}\langle\mathcal{O}(t)T_i(0)P_i\rangle$ computed numerically on the ensemble with $\kappa=0.14925$ and $\beta=1.75$ of the $V=24^3\times48$ lattice.}
\label{fig:correlators}
\end{figure}
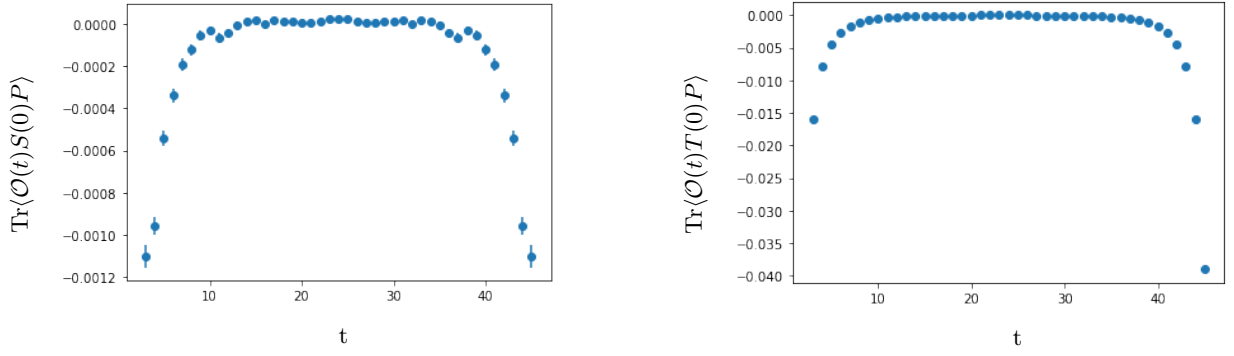

The spatial projectors $P_i=\gamma_4\gamma_i$ and $P_{ij}=\gamma_i \gamma_4 \gamma_j$ with $i=1,2,3$ are the most suitable to use numerically, as they lead to a solvable set of GIRS conditions (Eqs. (\ref{cond1_explicit}--\ref{cond4_explicit})) and a better signal-to-noise ratio. We plot the results for two of the correlators in Fig.~\ref{fig:correlators} to show the quality of the numerical signal. In order to alleviate the noise as much as possible, we use lattice symmetries to average different components of the correlators. Taking $\langle \mathcal{O}(t)S_i(0)P_i \rangle$ as an example (no sum over $i$), we can average all the different spatial directions $\mu,\nu=x,y,z$
\begin{equation}
    \langle \mathcal{O}(t)\overline{S}_i(0)P_i \rangle\rightarrow\frac{\langle \mathcal{O}(t)\overline{S}_x(0)P_x\rangle+\langle \mathcal{O}(t)\overline{S}_y(0)P_y\rangle+ \langle \mathcal{O}(t)\overline{S}_z(0)P_z\rangle }{3}.
\end{equation}
Secondly, because all the correlators are even under time reversal\footnote{This is opposite to the behavior of Eqs. (\ref{GSSMSbar}-\ref{GOOMSbar}), due to the anti-periodic boundary conditions.} one can also perform the average
\begin{equation}
\langle \mathcal{O}(t)S_i(0)P_i\rangle\rightarrow \frac{\langle \mathcal{O}(t)\overline{S}_i(0)P_i\rangle+\langle \mathcal{O}(N_t-t)\overline{S}_i(0)P_i\rangle}{2}.
\end{equation}
Finally due to charge conjugation Eq \eqref{GFsidentity} the following relation between Green's functions holds
\begin{equation}
    \text{Tr}\Big[\langle \mathcal{O}(t)\overline{S}_i(0)\rangle P_i\Big] = \text{Tr}\Big[C\langle S_i(0)\overline{\mathcal{O}}(t)\rangle^TC^{-1}P_i\Big]=  \text{Tr}\Big[\langle S_i(t)\overline{\mathcal{O}}(0)\rangle P_i\Big ], 
\end{equation}
so that one can also average the reverse ordered operators,
\begin{equation}
    \text{Tr}\Big[\langle \mathcal{O}(t)\overline{S}_i(0)\rangle P_i\Big]\rightarrow\frac{\text{Tr}\Big[\langle \mathcal{O}(t)\overline{S}_i(0)\rangle P_i\Big] + \text{Tr}\Big[\langle S_i(t)\overline{\mathcal{O}}(0)\rangle P_i\Big ]  }{2}.
\end{equation}
The same arguments hold for the rest of the Green's functions. The signal can improve substantially after the averaging procedure. Due to possible autocorrelation we have only considered every 8th configuration of the ensemble.

\subsection{Renormalization coefficients}
For our choice of spatial projectors and after applying a summation over $\vec{x}$, the GIRS conditions on the lattice take the following form
\bea
\frac{1}{3 L^3} \sum_{\vec{x},\vec{y}} \sum_i \ {\rm Tr} \left[G^{SS, \,\GIRS}_{ii}((\vec{x},t), (\vec{y},0)) \ \gamma_i\gamma_4\gamma_i \right] &=&  \frac{2 (N_c^2 - 1) t}{\pi^2 {|t|}^5}, \label{lattGIRS_cond1} \\
\frac{1}{3 L^3} \sum_{\vec{x},\vec{y}} \sum_i \ {\rm Tr} \left[G^{TT, \,\GIRS}_{ii}((\vec{x},t), (\vec{y},0)) \ \gamma_i\gamma_4\gamma_i \right] &=&  \frac{5 (N_c^2 - 1) t}{4 \pi^2 {|t|}^5}, \label{lattGIRS_cond2} \\
\frac{1}{3 L^3} \sum_{\vec{x},\vec{y}} \sum_i \ {\rm Tr} \left[G^{ST, \,\GIRS}_{ii}((\vec{x},t), (\vec{y},0)) \ \gamma_i\gamma_4\gamma_i \right] &=&  \frac{(N_c^2 - 1) t}{\pi^2 {|t|}^5}, \label{lattGIRS_cond3} \\
\frac{1}{3 L^3} \sum_{\vec{x},\vec{y}} \sum_i \ {\rm Tr} \left[G^{S{\cal O}, \,\GIRS}_{i}((\vec{x},t), (\vec{y},0)) \ \gamma_4 \gamma_i \right] &=& 0. \label{lattGIRS_cond4} 
\eea
The renormalized Green's functions appearing in Eqs.~(\ref{lattGIRS_cond1} -- \ref{lattGIRS_cond4}) are related to the bare Green's functions extracted from lattice simulations via Eqs.~(\ref{cond1_explicit} -- \ref{cond4_explicit}), where the Z factors stand for $Z^{{\rm L},{\rm GIRS}}$. 

\begin{figure}
    \centering
    \begin{subfigure}{1\textwidth}
    \begin{tikzpicture}
    \node (img)  {\includegraphics[width=.45\textwidth]{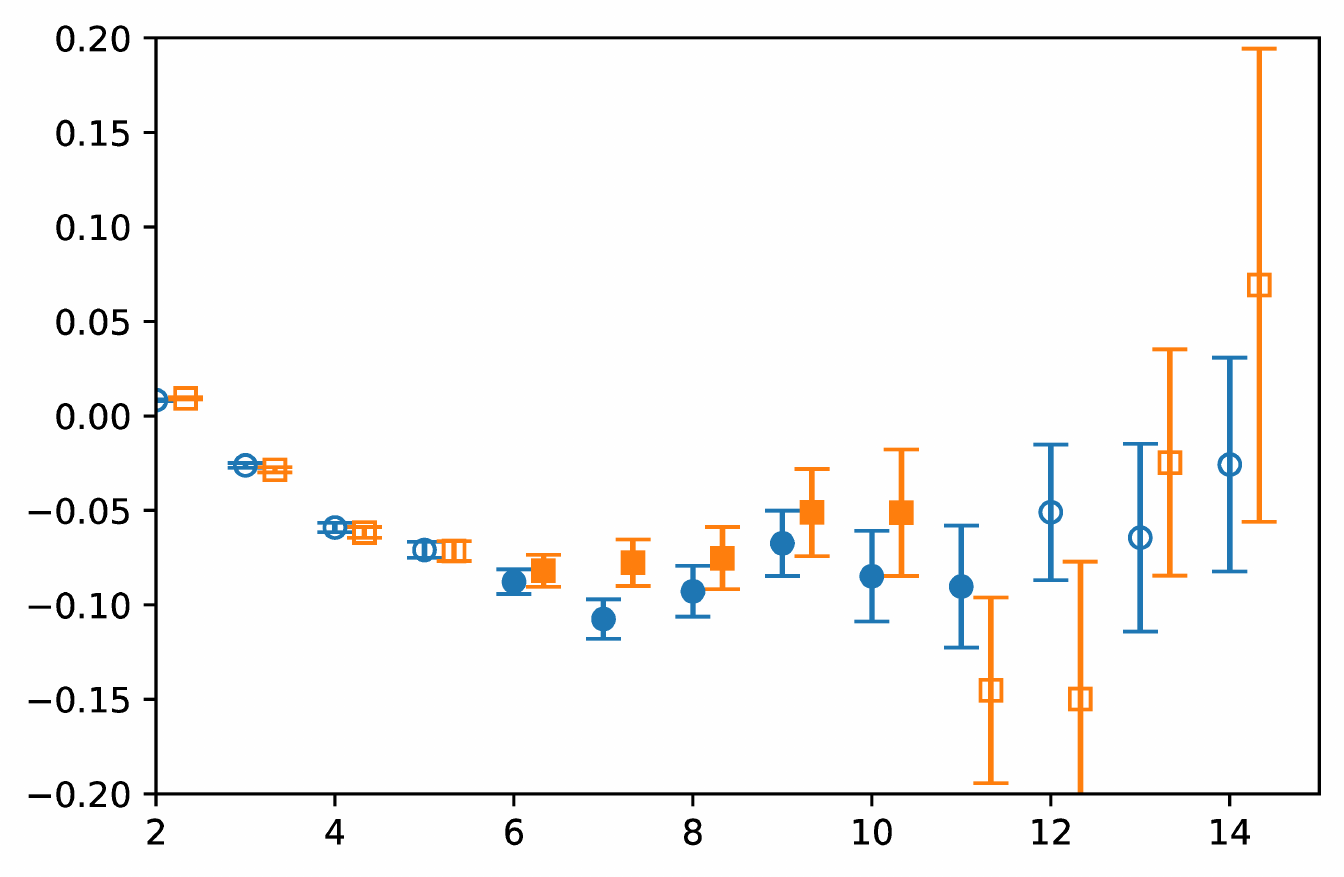}};
      \node[above,font=\bfseries] at (current bounding box.north) {GIRS scheme};
      \node[below=of img, node distance=0cm, xshift=0.5cm, yshift=1cm] {t};
      \node[left=of img, node distance=0cm, rotate=90, anchor=center,yshift=-0.7cm,] {$Z_{ST}/Z_{SS}$}; 
     \end{tikzpicture}
    \begin{tikzpicture}
    \node (img)  {\includegraphics[width=.45\textwidth]{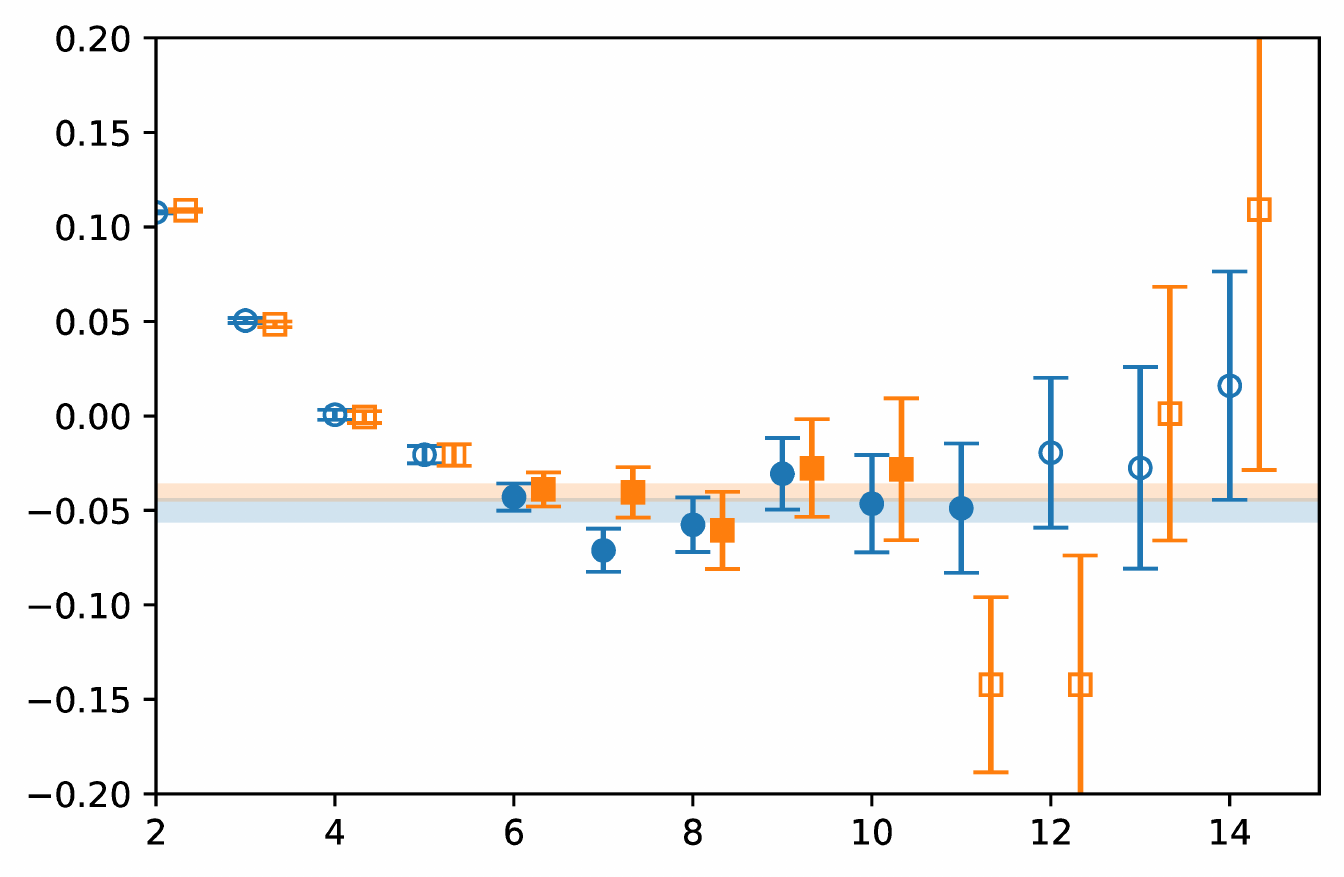}};
      \node[above,font=\bfseries] at (current bounding box.north) {$\MSbar$ scheme};
      \node[below=of img, node distance=0cm, xshift=0.5cm, yshift=1cm] {t};
      \end{tikzpicture}
    \centering
    \caption{ $V=24^3\times 48$ lattice with $\kappa=0.14925$ (blue dots) and $\kappa=0.14920$ (orange squares). The points on the $\kappa=0.14920$ ensemble are shifted in $t$ by $+0.33$ for visibility.}
    \label{24x48}
    \end{subfigure}
   \begin{subfigure}{1\textwidth}
    \begin{tikzpicture}
    \node (img)  {\includegraphics[width=.45\textwidth]{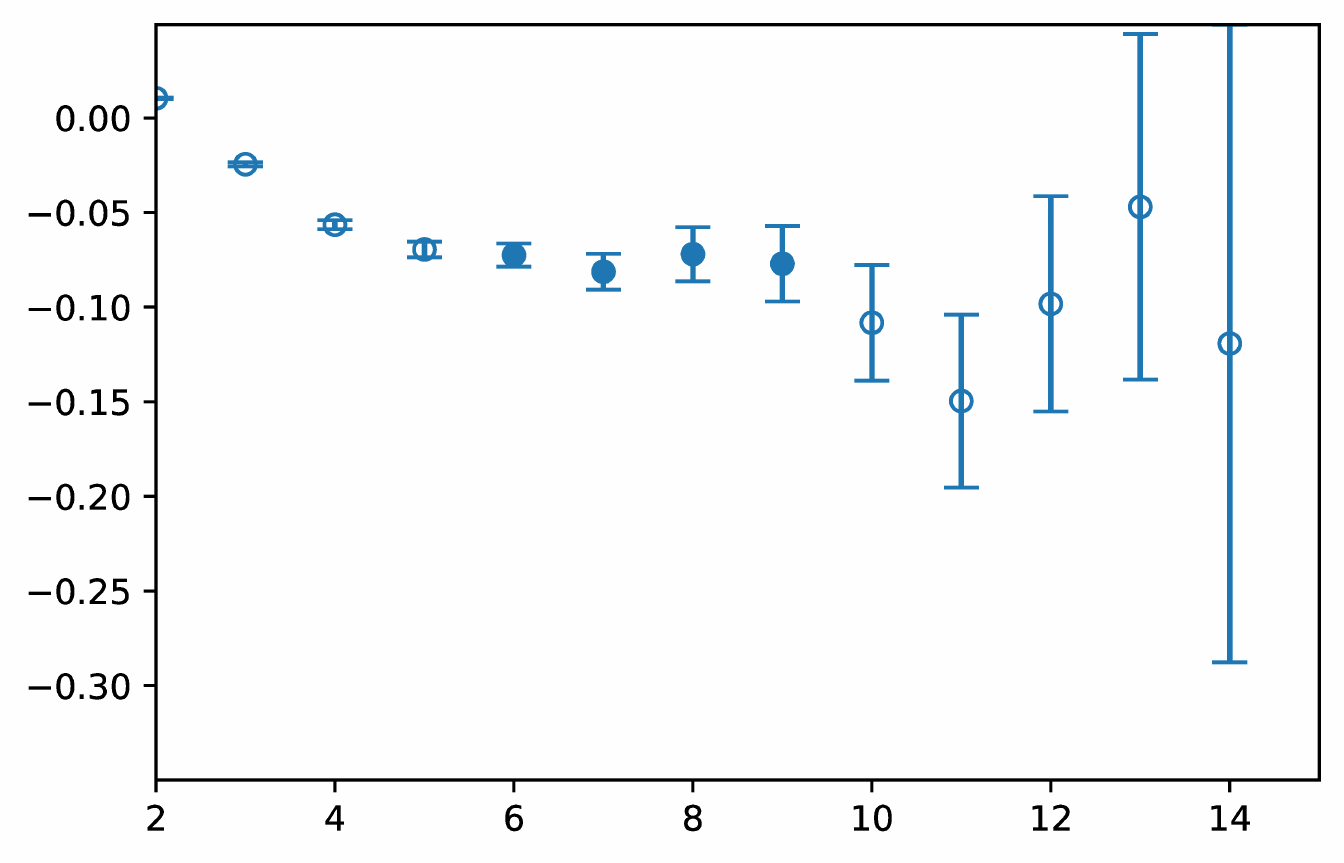}};
      \node[above,font=\bfseries] at (current bounding box.north) {GIRS scheme};
      \node[below=of img, node distance=0cm, xshift=0.5cm, yshift=1cm] {t};
      \node[left=of img, node distance=0cm, rotate=90, anchor=center,yshift=-0.7cm,] {$Z_{ST}/Z_{SS}$}; 
      \end{tikzpicture}
    \begin{tikzpicture}
    \node (img)  {\includegraphics[width=.45\textwidth]{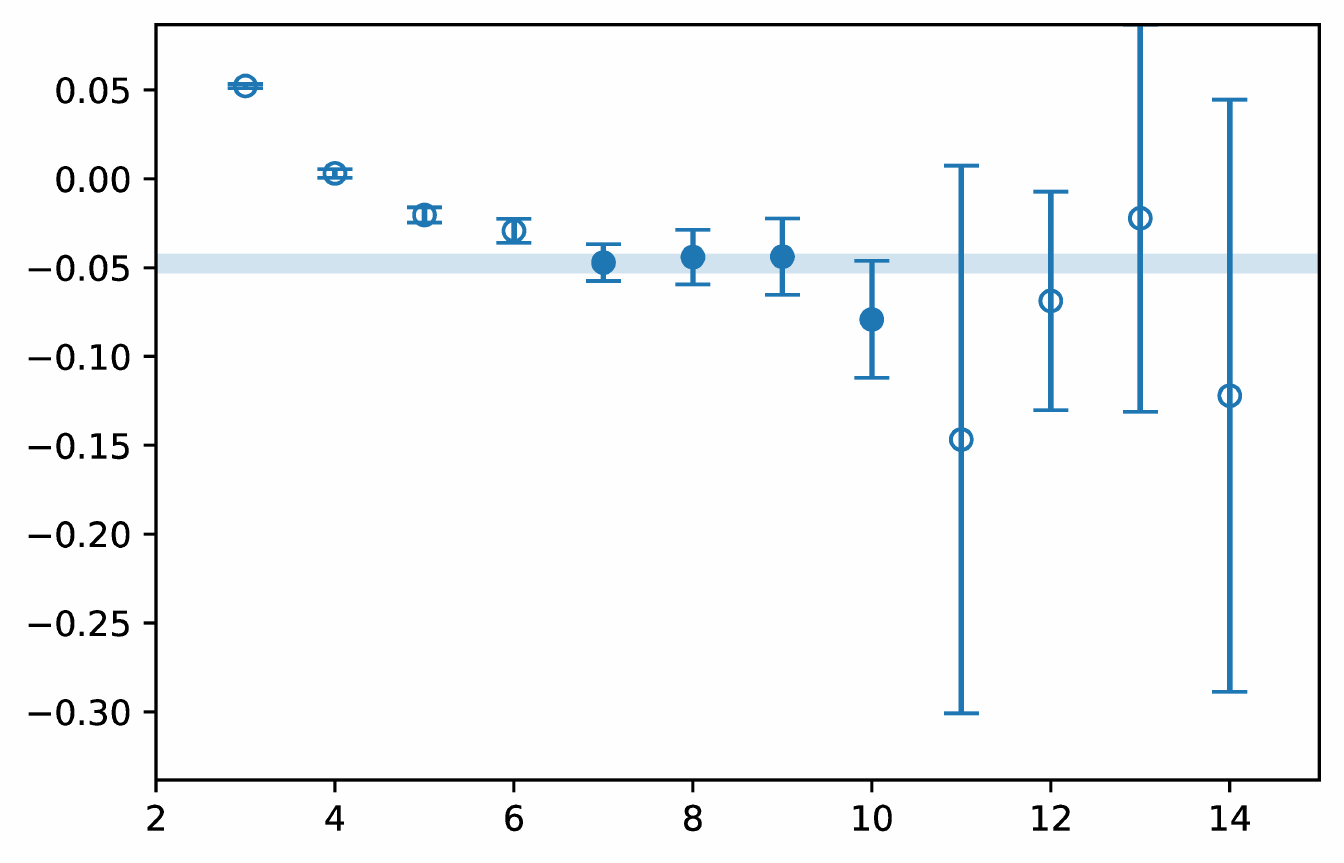}};
      \node[above,font=\bfseries] at (current bounding box.north) {$\MSbar$ scheme};
      \node[below=of img, node distance=0cm, xshift=0.5cm, yshift=1cm] {t};
      \end{tikzpicture}
    \caption{$V=32^3\times 64$ lattice with $\kappa=0.1494$.}
    \label{32x64}
    \end{subfigure}
       \begin{subfigure}{1\textwidth}
    \begin{tikzpicture}
    \node (img)  {\includegraphics[width=.45\textwidth]{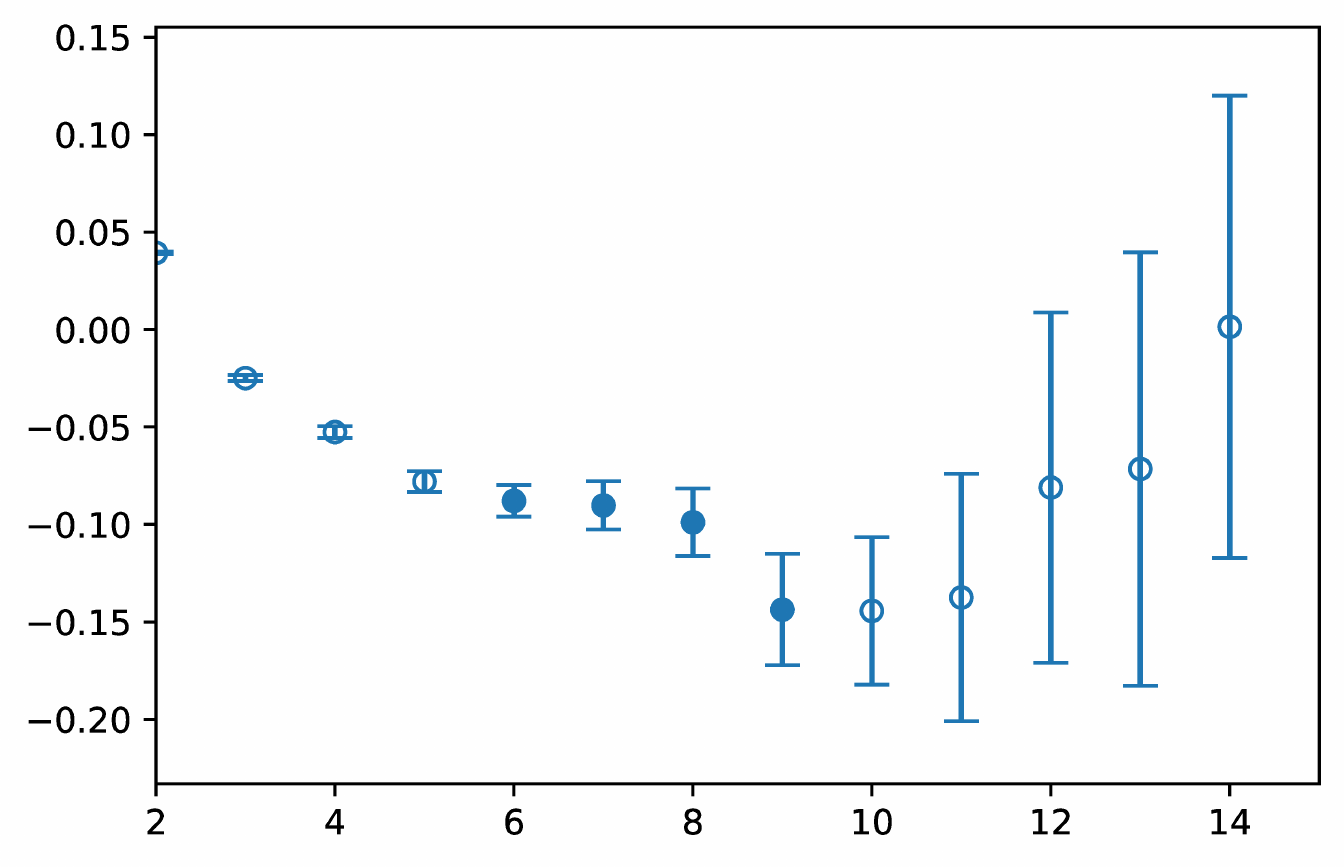}};
      \node[above,font=\bfseries] at (current bounding box.north) {GIRS scheme};
      \node[below=of img, node distance=0cm, xshift=0.5cm, yshift=1cm] {t};
      \node[left=of img, node distance=0cm, rotate=90, anchor=center,yshift=-0.7cm,] {$Z_{ST}/Z_{SS}$}; 
      \end{tikzpicture}
    \begin{tikzpicture}
    \node (img)  {\includegraphics[width=.45\textwidth]{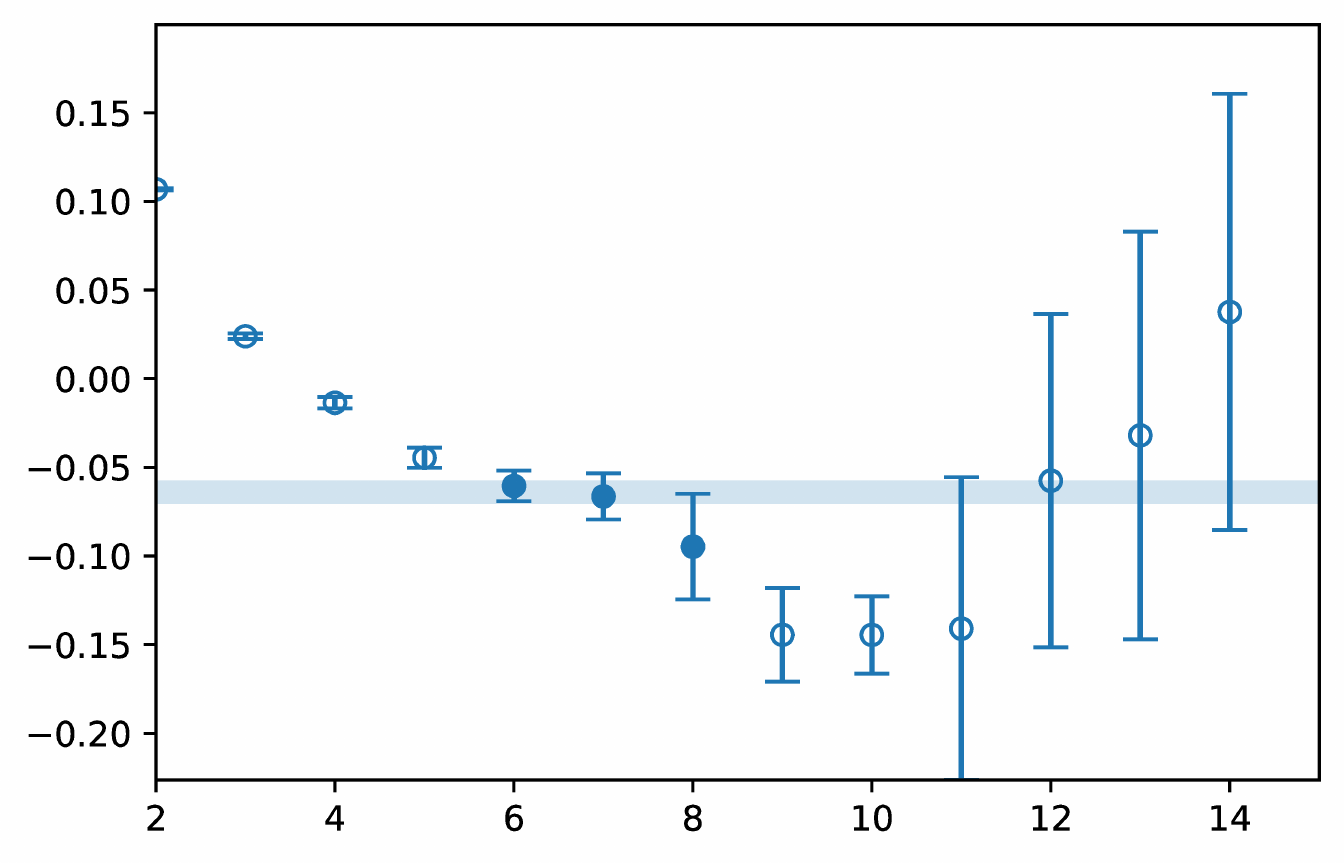}};
      \node[above,font=\bfseries] at (current bounding box.north) {$\MSbar$ scheme};
      \node[below=of img, node distance=0cm, xshift=0.5cm, yshift=1cm] {t};
      \end{tikzpicture}
    \caption{SU(3) ensemble on a $V=24^3\times 48$ lattice with $\kappa=0.1655$ and $\beta=5.6$.}
    \label{su3}
    \end{subfigure}
      \caption{$Z_{ST}/Z_{TT}$ in the GIRS (left) and in the $\MSbar$ scheme (right) as a function of time for the first time slices. The filled dots represent the regime where the effects of contact terms and noise are minor. The colour bands correspond to the linear extrapolation using the plateau interval.}
\label{fig:Z_factors}
\end{figure}
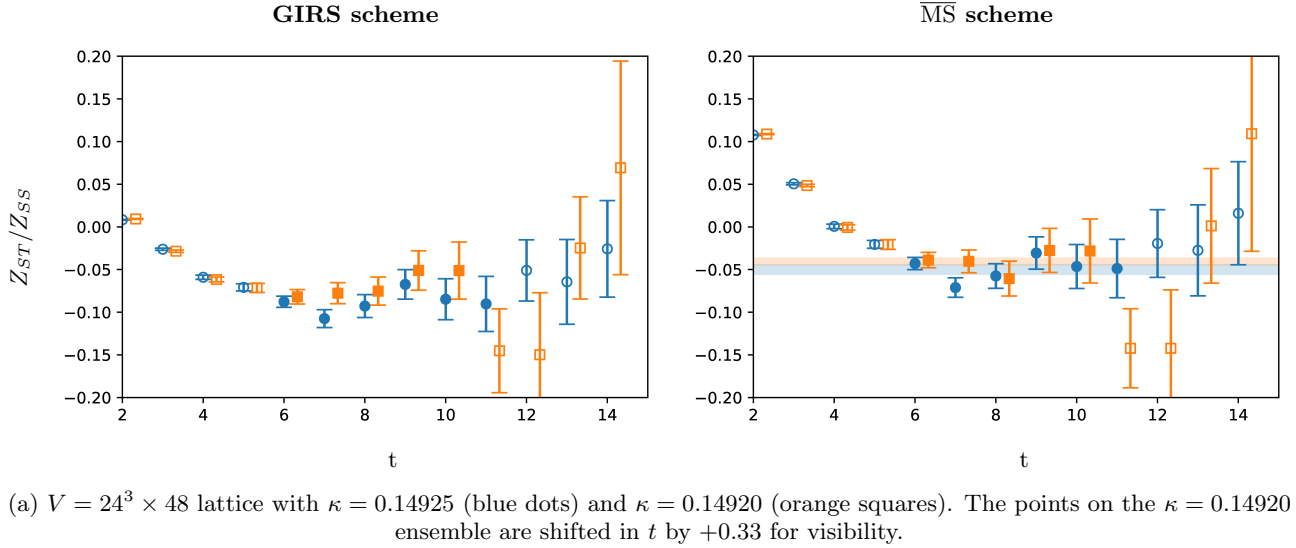
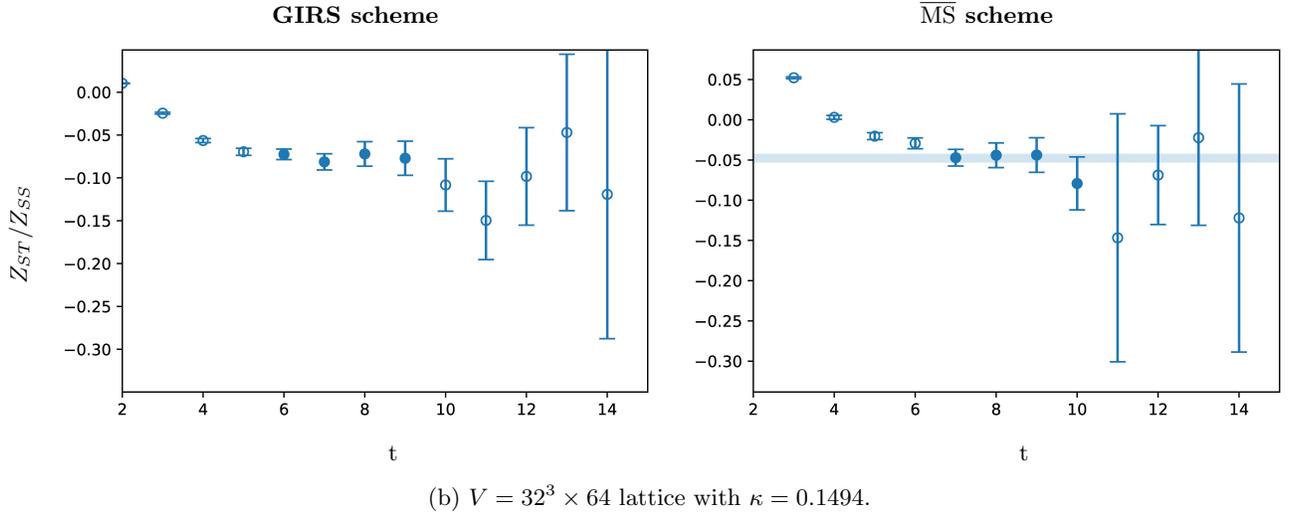
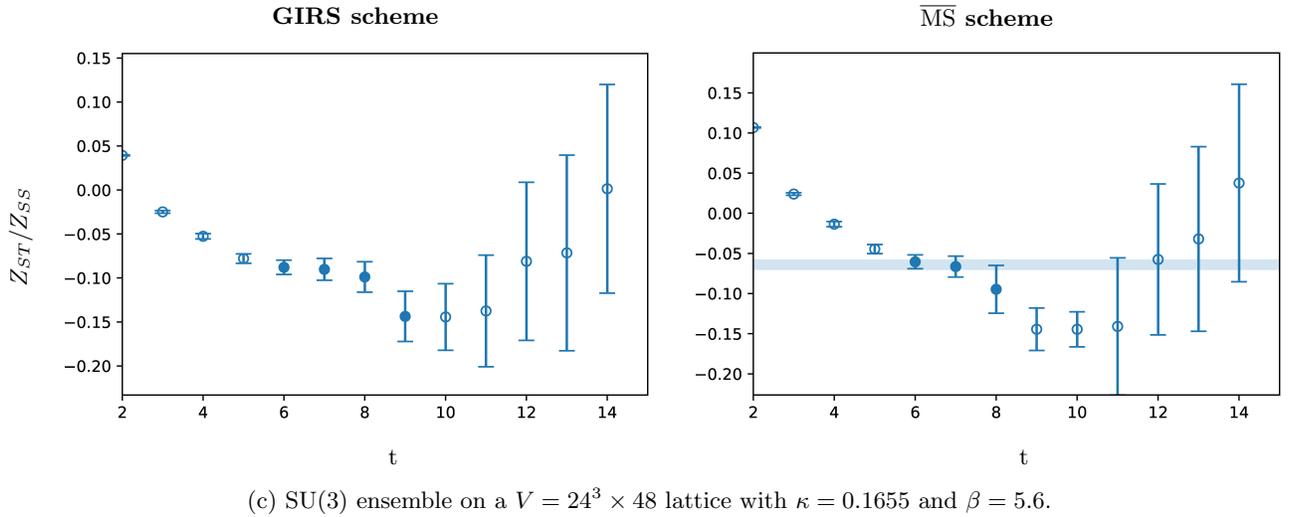

The GIRS renormalization conditions, Eqs. (\ref{cond1_explicit} -- \ref{cond4_explicit}), are enough to solve for all the renormalization factors of the supercurrents using the Green's functions computed on the lattice. The results for $Z_{ST}/Z_{SS}$ after Jackknife analysis are shown on the left side of Fig.~\ref{fig:Z_factors}. Due to gauge links extending in the temporal direction, short range effects persist to larger time separation $t$. On the other hand for large $t$ intervals $t>9$ the noise starts to become dominant and so the relevant signal is contained in only a rather small $t$ window (filled dots).


As already anticipated, all Z factors in GIRS can be translated to the $\MSbar$ scheme by applying the conversion factors  Eqs.~(\ref{convSS2}--\ref{convTT2}). All $\MSbar$ renormalization functions should turn out to be independent of the GIRS renormalization scale $t$ and a plateau-like behaviour of the Z factors in the $\MSbar$ scheme is expected. For instance, the ratio $Z_{ST}^{\LR,\MSbar} / Z_{SS}^{\LR,\MSbar}$ is extracted by combining the nonperturbative result for the GIRS mixing matrix $Z^{\LR,\GIRS}$ with the perturbative result for the conversion matrix $C^{\GIRS,\MSbar}$, through the following relation:
\be
\frac{Z_{ST}^{\LR,\MSbar}}{Z_{SS}^{\LR,\MSbar}} = \frac{C_{SS}^{\GIRS,\MSbar} Z_{ST}^{\LR,\GIRS} + C_{ST}^{\GIRS,\MSbar} Z_{TT}^{\LR,\GIRS}}{C_{SS}^{\GIRS,\MSbar} Z_{SS}^{\LR,\GIRS} + C_{ST}^{\GIRS,\MSbar} Z_{TS}^{\LR,\GIRS}}.
\ee
where we substituted the coupling constant $g^2_{\MSbar}$ on the conversion factors  Eqs.~(\ref{convSS2}--\ref{convTT2}) by the bare coupling constant of the lattice action, $\beta=2N_c/g^2$, since the difference would be $\mathcal{O} (g^4_{\MSbar})$. For comparison of the results, the lattice spacing can be estimated using the QCD Sommer scale value $r_0=0.5\text{ fm}$ which leads to a lattice spacing of $0.0554(11) \text{ fm}$ for SU($2$) and $0.0532(8)\text{ fm}$ for SU($3$) SYM.

The numerical values obtained for the ratio $Z_{ST}^{\LR,\MSbar} / Z_{SS}^{\LR,\MSbar}$ are presented on the right side of Fig.~\ref{fig:Z_factors}. The final values are determined by a constant fit in the plateau-like interval marked by filled dots. The results are collected in Table~\ref{tab:renormalization_factors}.
$Z_{ST}^{\LR,\MSbar} / Z_{SS}^{\LR,\MSbar}$ is much smaller than the perturbative predictions. Our perturbative estimate for SU($2$) SYM is $Z_{ST}^{\LR,\MSbar} / Z_{SS}^{\LR,\MSbar}=0.100809$, but compared to the nonperturbative simulations it does not include stout smearing. The value for SU($3$) SYM without clover improvement is $Z_{ST}^{\LR,\MSbar} / Z_{SS}^{\LR,\MSbar}=0.0656238$. In a consistent truncation at a given order in the coupling constant, one would use the tree level clover coefficient ($c_{\rm SW}=1.0$) and obtain $Z_{ST}^{\LR,\MSbar} / Z_{SS}^{\LR,\MSbar}=0.0508682$; with the one-loop improved value used in the actual simulations ($c_{\rm SW}=1.587$), one obtains  $Z_{ST}^{\LR,\MSbar} / Z_{SS}^{\LR,\MSbar}=0.0373759$.  The one-loop expressions for $Z_{ST}^{\LR,\MSbar}$ and $Z_{SS}^{\LR,\MSbar}$ are explicitly shown in Ref.~\cite{Bergner:2022wnb}, where the number of colors, $N_c$, the coupling constant $g$, and the clover parameter, $c_{\rm SW}$ are left unspecified.

\begin{figure}[ht]
    \centering
    \begin{subfigure}{1\textwidth}
    \begin{tikzpicture}
    \node (img) 
    {\includegraphics[width=.45\textwidth]{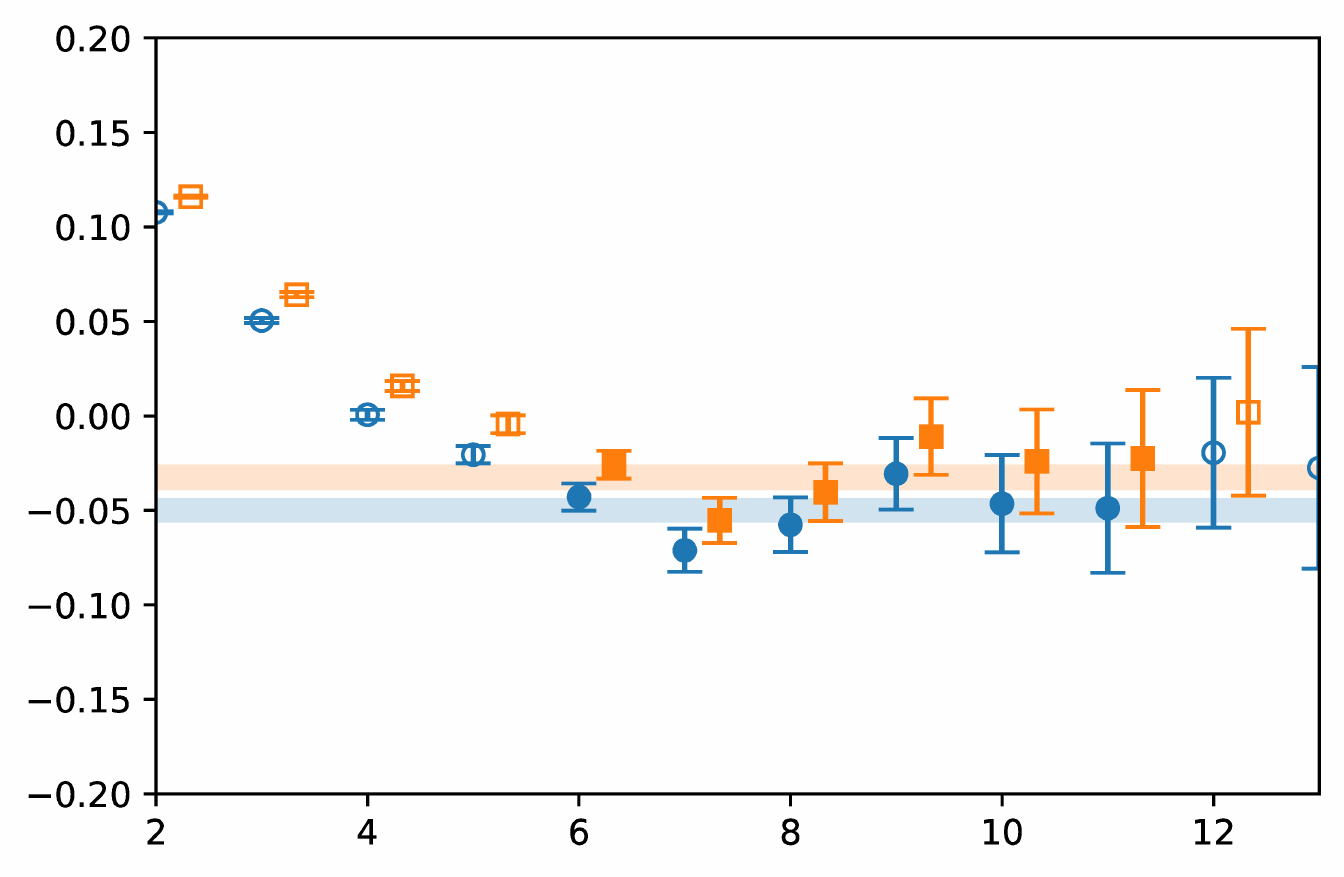}};
      \node[above,font=\bfseries] at (current bounding box.north) {$\MSbar$ scheme};
      \node[below=of img, node distance=0cm, xshift=0.5cm, yshift=1cm] {t};
      \node[left=of img, node distance=0cm, rotate=90, anchor=center,yshift=-0.7cm,] {$Z_{ST}/Z_{SS}$}; 
     \end{tikzpicture}
     \end{subfigure}
     \caption{$Z_{ST}^\MSbar/Z_{SS}^\MSbar$ computed with $C(g^2)^{\GIRS,\MSbar}$ (blue dots) and $C^{-1}(-g^2)^{\GIRS,\MSbar}$ (orange squares). The difference between the two computations $\Delta(Z_{ST}^\MSbar/Z_{SS}^\MSbar)=0.0277$ is due to higher loop corrections.}
     \label{fig:C_inv}
\end{figure}

Nonperturbative effects should be relevant and our parameter range is most likely far outside the perturbative regime where the one-loop computations can be reliable. In order to provide a further illustration for the importance of higher perturbative or nonperturbative corrections, we derived an alternative determination of $Z_{ST}^{\LR,\MSbar} / Z_{SS}^{\LR,\MSbar}$ using the substitution on the conversion matrix $C(g^2)^{\GIRS,\MSbar}\rightarrow C^{-1}(-g^2)^{\GIRS,\MSbar}$. At one-loop order this amounts to taking the inverse twice and so the result should be left invariant. Due to higher order corrections, this equivalence is, however, significantly violated as shown in Fig.~\ref{fig:C_inv}.

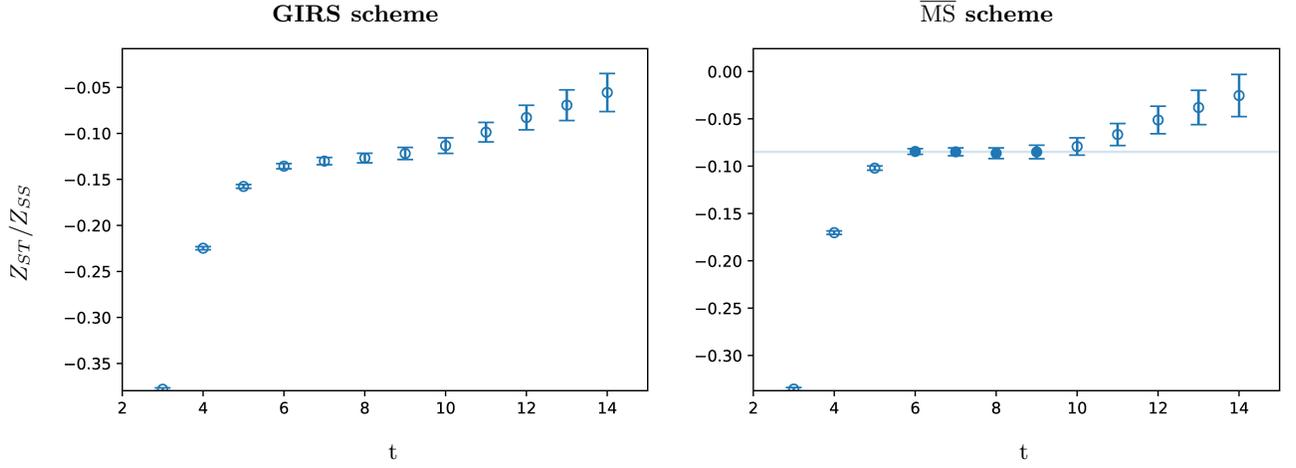
\begin{figure}[ht]
    \centering
    \begin{subfigure}{1\textwidth}
    \begin{tikzpicture}
    \node (img)  {\includegraphics[width=.45\textwidth]{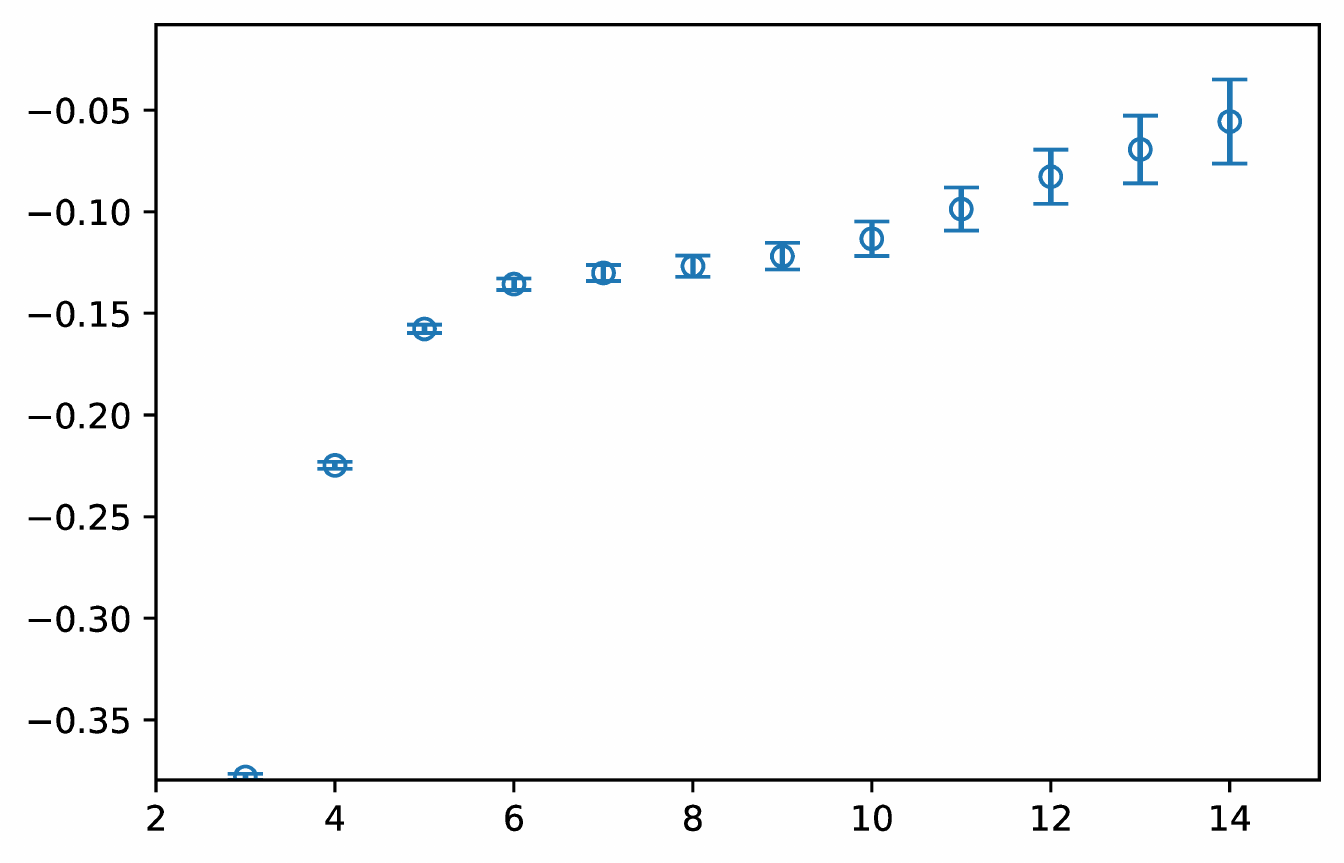}};
      \node[above,font=\bfseries] at (current bounding box.north) {GIRS scheme};
      \node[below=of img, node distance=0cm, xshift=0.5cm, yshift=1cm] {t};
      \node[left=of img, node distance=0cm, rotate=90, anchor=center,yshift=-0.7cm,] {$Z_{ST}/Z_{SS}$}; 
     \end{tikzpicture}
    \begin{tikzpicture}
    \node (img)  {\includegraphics[width=.45\textwidth]{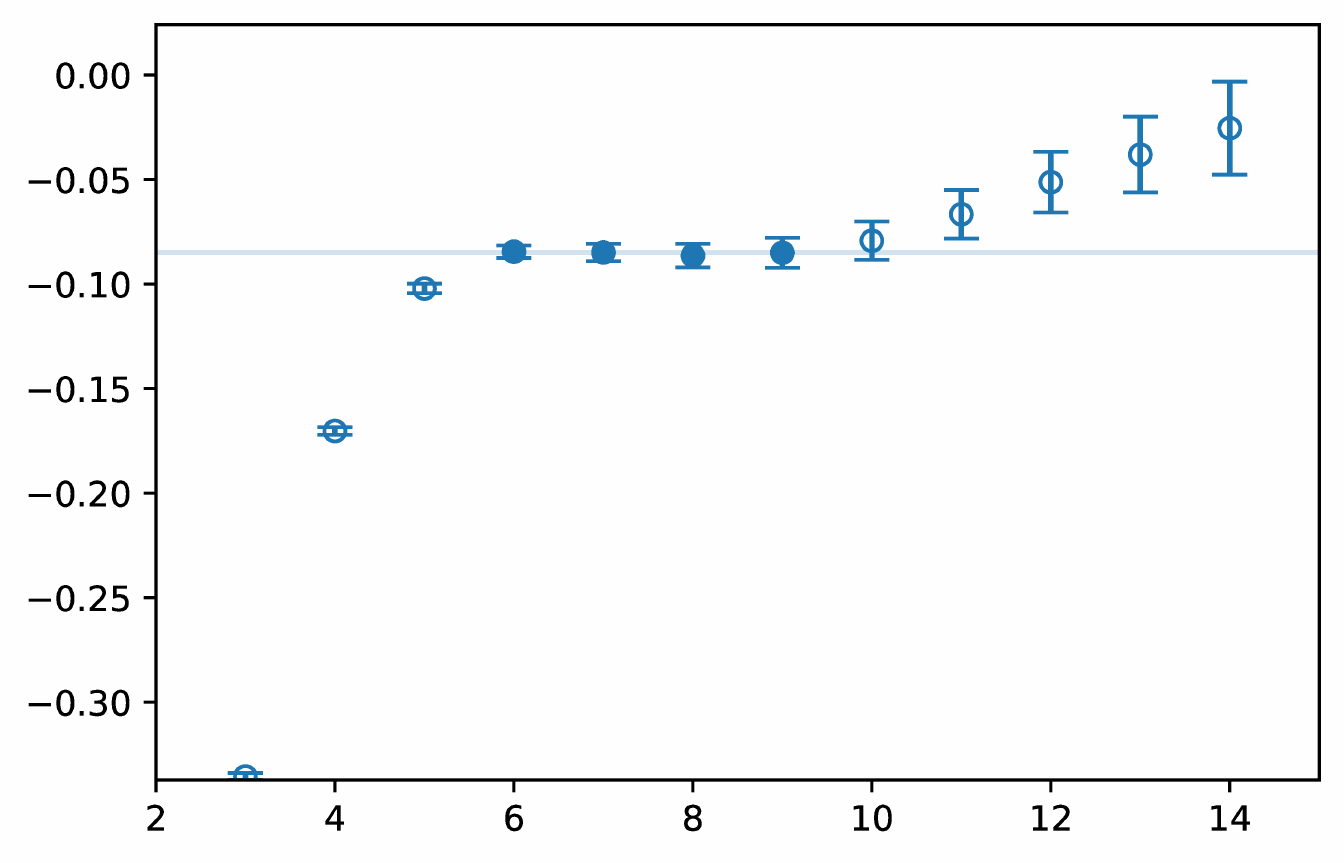}};
      \node[above,font=\bfseries] at (current bounding box.north) {$\MSbar$ scheme};
      \node[below=of img, node distance=0cm, xshift=0.5cm, yshift=1cm] {t};
      \end{tikzpicture}
    \centering
    \end{subfigure}
    \caption{ $V=24^3\times 48$ lattice at $\kappa=0.14920$ with smearing. The fitted value corresponds to $Z_{ST}/Z_{SS}= -0.0848(15)$. A more pronounced plateau is observed in the $\MSbar$ data, as expected.}
    \label{smeared}
\end{figure}

In order to reduce higher loop corrections one would ideally like to simulate at smaller couplings close to the perturbative regime, however the cost of such simulations can increase rather fast. As an exploratory and alternative approach we have done measurements on smeared configurations to investigate how much the noise is reduced and whether the results get closer to the perturbative predictions. We applied 6 levels of stout smearing with smearing parameter $\rho=0.15$ and adjusted the mass parameter such that the pion mass in lattice units stays approximately the same. The resulting ratio $Z_{ST}/Z_{SS}$ is presented in Fig.~\ref{smeared} both in GIRS and in the $\MSbar$ scheme. We can observe that the noise is drastically decreased and a plateau-like intermediate $t$ range becomes much more visible. In GIRS scheme we observe a clear $t$ dependence which is erased in the $\MSbar$ scheme by the conversion factors up to a certain $t$, where higher order corrections of the conversion matrix seem to become relevant. We note that the fitted value has now moved away even further from the perturbative result; however, a faithful comparison would require performing perturbative computation with smeared gauge links, which we have not done so far.


\begin{table}[h!]
    \centering
    \begin{tabular}{|c|c|c|c|c|c|c|c|c|c|c|c|}
        \hline
       Gauge group & $\beta$ & $\kappa$ & $w_{0,\chi} m_\pi$ & Scheme & L & T & $Z_{SS}$ & $Z_{ST}$ & $Z_{TS}$ & $Z_{TT}$ & $Z_{ST}/Z_{SS}$\\
        \hline
        SU($2$) & 1.75  &0.14920& 0.6915(62) & $\MSbar$& 24 & 48 & ~0.745(20)~&-0.0362(48) & 0.1759(12) & 0.3518(30) & -0.0499(55)  \\
        \hline
        SU($2$) & 1.75 &0.14925 & 0.6467(92) & $\MSbar$ & 24 & 48 & 0.783(24) & -0.0312(26) & 0.201(16) & 0.3612(67) &  -0.0404(38) \\
        \hline
        SU($2$) & 1.75 &0.14940 & 0.5471(80)& $\MSbar$ & 48 & 64 & 0.877(25) & -0.0417(38) & 0.275(20) & 0.275(20) & -0.0477(44) \\
        \hline
        SU($3$) & 5.6 &0.16550 & 1.204(27) & $\MSbar$ & 24 & 48 & 1.127(43)  & -0.0703(56)  & 0.323(20) & 0.4302(25)  & -0.0640(55)\\
        \hline
        
    \end{tabular}
    \caption{Value of the Z renormalization factors of the supercurrents obtained nonperturbatively on the GIRS and then translated to the $\MSbar$ scheme by using the conversion matrix $C^{\GIRS,\MSbar}$. The errors presented here are purely statistical, resulting from a Jacknife analysis. For reference, the parameters of the different ensembles are also provided. Each simulation has been done on an $L^3\times T$ lattice and the adjoint pion mass is provided in units of the gradient flow scale $w_{0,\chi}$ extrapolated to the chiral limit ($w_{0,\chi}/a=3.411(18)$ for SU(2) SYM and $w_{0,\chi}/a=3.485(71)$ for SU(3) SYM).}
    \label{tab:renormalization_factors}
\end{table}

\begin{figure}[ht]
\centering 
\includegraphics[scale=1]{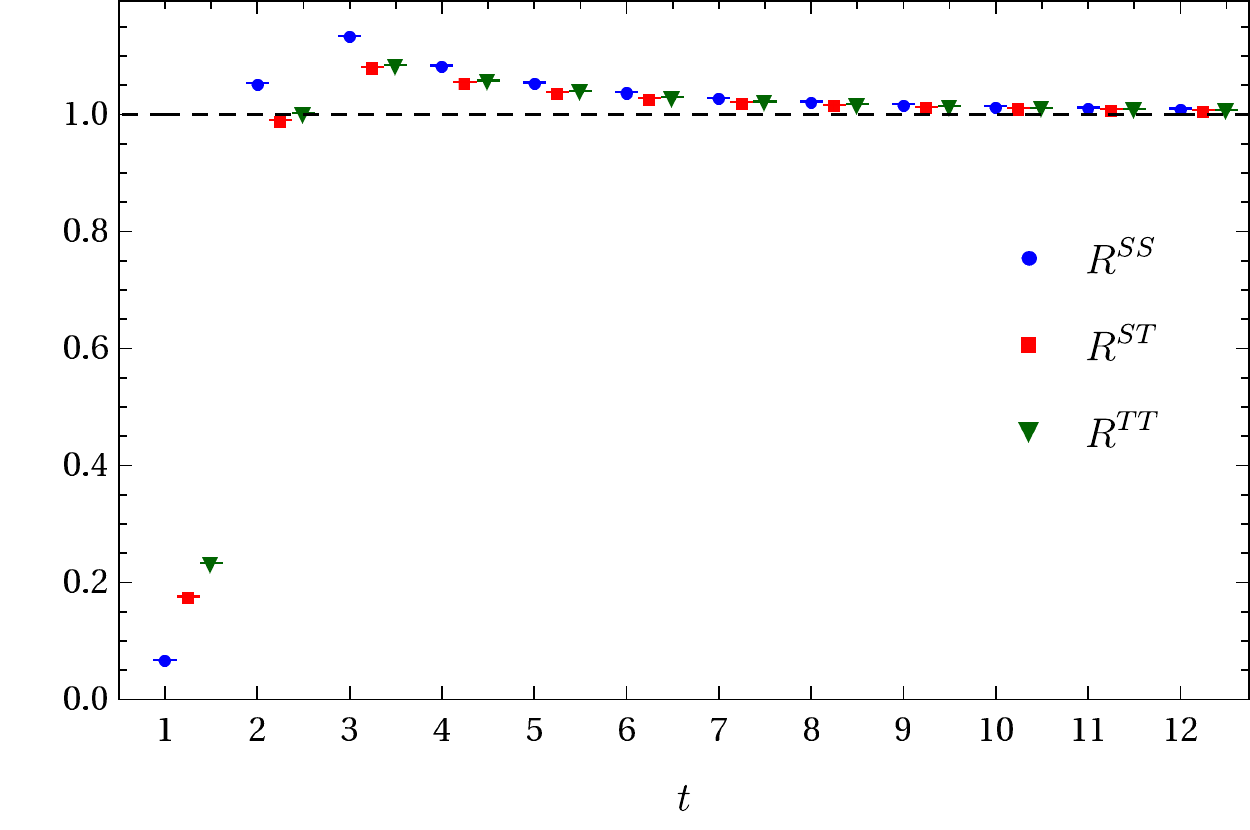}
\caption{Ratios between lattice (all orders in $a$) and continuum tree-level values: $R^{AB} (t) \equiv {\rm Tr} [{(G_{ii}^{AB} (t))}^{\rm tree \ lat.} \gamma_i \gamma_4 \gamma_i] / {\rm Tr} [{(G_{ii}^{AB} (t))}^{\rm tree \ cont.} \gamma_i \gamma_4 \gamma_i]$, for $AB = SS, ST, TT$. \protect\linebreak For better visibility, $R^{ST} (t)$ and $R^{TT} (t)$ are shifted in $t$ by +0.25 and +0.50, respectively.}
\label{artifacts_ratio}
\end{figure}

For improving the estimates, one may employ a tree-level correction by subtracting tree-level discretization effects from the correlators. This can be useful especially for small values of $t$, where large cut-off effects are present. An equivalent procedure (up to higher-order effects) is to replace the continuum tree-level values in the r.h.s. of Eqs.~(\ref{lattGIRS_cond1} -- \ref{lattGIRS_cond4}) with the corresponding lattice tree-level values, which include artifacts to all orders in the lattice spacing $a$. We examine the effect of this tree-level correction by calculating the ratio between the lattice and continuum tree-level values: $R^{AB} (t) \equiv {\rm Tr} [{(G_{ii}^{AB} (t))}^{\rm tree \ lat.} \gamma_i \gamma_4 \gamma_i] / {\rm Tr} [{(G_{ii}^{AB} (t))}^{\rm tree \ cont.} \gamma_i \gamma_4 \gamma_i]$, where $A,B = S,T$. In Fig.~\ref{artifacts_ratio}, we show the ratios $R^{SS} (t), R^{ST} (t)$ and $R^{TT} (t)$. We observe that the most important differences between $\mathcal{O} (a^0)$ and all orders in $a$ corrections of tree-level values regard very small values of $t$, for which results in $Z_{ST}/Z_{SS}$ are far from a plateau; thus these differences do not significantly affect the plateau values.

\newpage

\section{Discussion}
\label{Discussion}

In this paper, we have presented a concrete prescription to renormalize nonperturbatively the supercurrent operator in $\mathcal{N}=1$ Supersymmetric Yang-Mills (SYM) theory on the lattice, using a gauge-invariant scheme (GIRS). The employed scheme addresses the mixing between the supercurrent and all gauge-invariant operators which respect the same global symmetries and share the same quantum numbers. An advantageous feature of GIRS is that whole classes of gauge-noninvariant operators, which, {\it in principle}, can and will mix with the supercurrent, have zero contributions to the gauge-invariant Green's functions of GIRS. In this respect, the set of mixing operators that one needs to study within GIRS is greatly reduced. This makes GIRS a good alternative for renormalizing operators in the presence of mixing, compared to more standard methods, such as employing Ward Identities on the lattice.

We have employed GIRS on two lattices of SU($2$) SYM with three different values of critical mass parameter, on a smeared ensemble, and on an ensemble based on the gauge group SU($3$). We have presented nonperturbative results for the mixing matrix. In parallel, we have performed a one-loop perturbative calculation of the mixing matrix in both GIRS and $\MSbar$ schemes in order to deduce the conversion matrix between the two schemes. At the end, we have combined the nonperturbative data with the one-loop conversion matrix and we have extracted the $\MSbar$ mixing matrix. Our result for the ratio $Z_{ST}^{\LR,\MSbar} / Z_{SS}^{\LR,\MSbar}$ is different from our previous perturbative computation at one loop \cite{Bergner:2022wnb}. However, the parameter regime for our first test is most likely far outside the perturbative regime.

One means of improving the nonperturbative estimates is further elimination of the discretization errors. For small values of $t$, large cutoff  effects are present in our calculation. In order to address this issue, a higher-loop perturbative evaluation of lattice artifacts, to all orders in the lattice spacing, for different values of $t$, could be performed. As has been observed in other contexts \cite{Capitani:2000xi,Constantinou:2009tr,Gockeler:2010yr,Constantinou:2013ada,Alexandrou:2015sea,Constantinou:2022aij}, subtraction of the unwanted contributions of the finite lattice spacing from the nonperturbative estimates can lead to a more rapid convergence to the continuum limit. Such a procedure has been successfully employed to Green's functions of operators in momentum space up to one loop. However, in a coordinate-space scheme, such as GIRS, a one-loop computation involves two-loop Feynman diagrams, which make the application of this subtraction in GIRS more difficult.  

A second improvement is the elimination of truncation effects coming from the conversion matrix. For large values of $t$, the one-loop conversion matrix is not enough to guarantee convergence, since strong higher-loop logarithmic divergences in $t$ are present. A natural continuation of the present work is the two-loop calculation of the conversion matrix.

From the numerical point of view, more statistics, especially for larger values of $t$, can enlarge the window of time that can be used for reaching the plateau in the renormalization factors. Finally simulating closer to a perturbative regime would be a very useful way to see if one can match the perturbative with the nonperturbative result.


\begin{acknowledgments}
 We thank G.~Münster for helpful discussions and comments.
 G.~Bergner\ and I.~Soler\ acknowledge financial support from the Deutsche Forschungsgemeinschaft (DFG) Grant No.~BE 5942/3-1 and 5942/4-1. 
 This work was co-funded by the European Regional Development Fund and the Republic of Cyprus through the Research and Innovation Foundation (Projects: EXCELLENCE/0918/0066 and EXCELLENCE/0421/0025). 
 M.~Costa also acknowledges partial support from the Cyprus University of Technology under the ``POST-DOCTORAL" programme. G.~Spanoudes\ acknowledges financial support from H2020 project PRACE-6IP (Grant agreement ID: 823767). The authors gratefully acknowledge the Gauss Centre for Supercomputing e.~V.\ (www.gauss-centre.eu) for funding this project by providing computing time on the GCS Supercomputer SuperMUC-NG at Leibniz Supercomputing Centre (www.lrz.de). Further computing time has been provided on the compute cluster PALMA
 of the University of M\"unster.
\end{acknowledgments}

\bibliography{SuperCurrent.bib}

\end{document}